\documentclass[sigconf,9pt]{acmart}

\usepackage[english]{babel}
\usepackage{cleveref}
\crefname{section}{}{\S\S}
\crefdefaultlabelformat{\S#2#1#3}
\usepackage{xspace}
\usepackage{comment}
\usepackage[normalem]{ulem}
\usepackage{enumitem}
\setlist{nolistsep}
\usepackage{listings}
\usepackage{color}
\definecolor{dkgreen}{rgb}{0,0.6,0}
\definecolor{gray}{rgb}{0.5,0.5,0.5}
\definecolor{mauve}{rgb}{0.58,0,0.82}
\definecolor{agentblue}{RGB}{70, 130, 180}
\definecolor{memorygrey}{RGB}{169, 169, 169}
\definecolor{dockerorange}{RGB}{255, 140, 0}
\definecolor{llmpurple}{RGB}{147, 112, 219}
\usepackage{algorithm}
\usepackage{algpseudocode}   

\usepackage{soul}

\usepackage[font=footnotesize]{caption}
\graphicspath{{figures/}}
\usepackage{graphicx}
\usepackage{multirow}
\usepackage[utf8]{inputenc}
\usepackage{tikz}
\usetikzlibrary{shapes.geometric, shapes.symbols, arrows.meta, positioning, calc, shadows}
\usepackage{pifont}
\usepackage{subcaption}
\usepackage{graphicx}
\usepackage{makecell}

\newcommand{\cmark}{\ding{51}}
\newcommand{\xmark}{\ding{55}}
\newcommand{\sys}{\emph{RepLLM}\xspace} 

\newcommand{\hhy}[1]{{\color{black} #1}}
\newcommand{\xyx}[1]{{\color{black} #1}}

\acmYear{2026}\copyrightyear{2026}
\setcopyright{cc}
\setcctype[4.0]{by}
\acmConference[SIGCOMM '26]{ACM SIGCOMM 2026 Conference}{August 17--21, 2026}{Denver, CO, USA}
\acmBooktitle{ACM SIGCOMM 2026 Conference (SIGCOMM '26), August 17--21, 2026, Denver, CO, USA}
\acmDOI{10.1145/3789240.3829170}
\acmISBN{979-8-4007-2467-1/26/08}

\acmPrice{}

\begin{document}

\title{\sys: Toward Automatically Reproducing Network Research Results}
\renewcommand{\shorttitle}{\sys: Toward Automatically Reproducing \\ Network Research Results}



\author{ 
{\large
Yining Jiang\textsuperscript{1}, 
Yunxin Xu\textsuperscript{1}, 
Wenyun Xu\textsuperscript{1}, 
Yufan Zhu\textsuperscript{2}, 
Rui Liu\textsuperscript{1}, 
Tangtang He\textsuperscript{2}, 
Haiying Huang\textsuperscript{1}, 
Letian Zhu\textsuperscript{1}, 
Qingyu Song\textsuperscript{1}\footnotemark[1], 
Qiang Su\textsuperscript{1}, 
Lizhao You\textsuperscript{1}, 
Lu Tang\textsuperscript{1}, 
Wanjian Feng\textsuperscript{3}, 
Yuchao Zhang\textsuperscript{4}, 
Linghe Kong\textsuperscript{5}, 
Qiao Xiang\textsuperscript{1}\footnotemark[1], 
Jiwu Shu\textsuperscript{1,6} 
\\[0.5em]
{\itshape
\textsuperscript{1}Fujian Engineering Research Center of High-Performance Intelligent Computing Systems, School of Informatics, Xiamen University \quad
\textsuperscript{2}Xiamen University \quad 
\textsuperscript{3}Yealink \quad 
\textsuperscript{4}Beijing University of Posts and Telecommunications \\ 
\textsuperscript{5}Shanghai Jiaotong University \quad 
\textsuperscript{6}Tsinghua University
}
}
}

\makeatletter
\gdef\authors{%
Yining Jiang, Yunxin Xu, Wenyun Xu, Yufan Zhu, Rui Liu, 
Tangtang He, Haiying Huang, Letian Zhu, Qingyu Song, Qiang Su, Lizhao You, Lu Tang,
Wanjian Feng, Yuchao Zhang, Linghe Kong, Qiao Xiang, Jiwu Shu%
}
\makeatother
\renewcommand{\shortauthors}{Jiang et al.}

\begin{abstract}
Result reproduction of computer networking research is challenging as the scarcity of open-source implementations and the complexity of heterogeneous system architectures. Even though Large Language Models have demonstrated potential in code generation, existing code generation frameworks often fail to address the long-context constraints and intricate logical dependencies, which are vital in reproducing network systems from academic papers. 
Thus, we introduce \emph{RepLLM}, an end-to-end multi-agent framework designed to automate code reproduction from paper content. \emph{RepLLM} features a collaborative architecture comprising four specialized agents---Content Parsing, Architecture Design, Code Generation, and Audit \& Repair, which are coordinated through \textit{Shared Memory} mechanism to ensure global context consistency. 
With the enhancement of Structured Chain-of-Thought LLM reasoning and a sandbox-isolated static-dynamic debugging methodology, our framework effectively resolves semantic discrepancies and runtime errors, thereby improving reliable reproductions. Extensive evaluations on representative papers in top conferences demonstrate that \emph{RepLLM} outperforms state-of-the-art system-level LLM frameworks in generating compile-ready and logically correct systems. Our results show that, with the aid of \emph{RepLLM}, we can reproduce 95\% of the original benchmarks within approximately two hours while reducing token consumption by up to 10\% compared with state-of-the-art baselines.
\end{abstract}



\maketitle
\footnotetext[1]{Qingyu Song and Qiao Xiang are corresponding authors.}

\section{Introduction}
\label{sec:introduction}

Reproducibility is considered as the cornerstone of computer networking research, offering substantial value to both the academic and industrial communities. Beyond ensuring the reliability and validity of research findings, reproducible results provide the industry with a rigorous foundation for technology selection and system optimization~\cite{handigol2012reproducible}. 

Result reproduction is challenging due to the limited availability of artifacts and the complexity of re-implementation. Current reproduction strategies generally fall into two categories: utilizing executable prototypes provided by authors~\cite{yang2015real} or third parties~\cite{zhang2020apkeep}, or undertaking implementation on the textual description of the original paper. However, most artifacts in networking are unavailable. Analysis reveals that even within top conferences such as SIGCOMM and NSDI, only 38.05\% of articles published over the past decade include open-source implementations. Thus, researchers are often forced to rely on re-implementation from scratch. 
However, re-implementation of a network system is resource-intensive and requires a strong command of domain-specific knowledge. This high barrier to entry stands in sharp contrast to adjacent fields like machine learning, where standardized frameworks often allow results to be replicated in a few days~\cite{yan2017learning}. 

\hhy{To reduce reproduction costs, recent research has increasingly turned to Large Language Models (LLMs) for automated code generation, based on the proficiency in completion, debugging, and synthesis~\cite{github_copilot,openai_chatgpt,zhang2024pydex,rahmani2021multi,yen2021semi}.} \hhy{Notable progress has been made in ``paper-to-code'' generation~\cite{seo2025paper2code}, particularly within the machine learning (ML)~\cite{lin2025autop2c} and biomedical~\cite{luo2025intention} fields.}
\hhy{However, existing LLM-based methods remain largely domain-specific and are ill-suited for re-implementing code in network systems research. Network research spans diverse areas, including protocols, algorithms, and system architectures, and lacks a unified programming paradigm. For example, systems such as NetChain\cite{jin2018netchain}, GRoot\cite{kakarla2020otot}, Teal\cite{xu2023teal} differ in their architectures. By contrast, existing ML code generation approaches typically benefit from standardized programming frameworks, such as PyTorch~\cite{paszke2019pytorch} and TensorFlow~\cite{abadi2016tensorflow}.}

\hhy{While recent approaches have attempted to bridge this gap, they remain limited to specific scenarios in which code generation tasks are defined by well-structured inputs and outputs. For example, some studies focus on translating structured specifications, such as RFCs or configurations, into code~\cite{chen2022software,yen2021semi}, which require structured inputs that are rarely available in broader systems research. \citet{wang2025llm} restrict their scope to satellite networks. In addition, \citet{xiang2023toward} reproduce the code for four papers spanning diverse areas. However, their approach relies on domain-specific prompts that require exhaustive human interactions.}

\hhy{The re-implementation of network research code with LLMs remains challenging. First, networking papers are typically lengthy and difficult for LLMs to process comprehensively. Providing excessive context at once can degrade the performance of large models~\cite{liu2023repobench,shi2023large}. Second, generating complex network codebases with LLMs remains difficult. Existing studies~\cite{jimenez2023swe,zelikman2023parsel,bairi2024codeplan} show that generating complex, multi-file codebases in a single inference pass is beyond the capabilities of current LLMs. Network research often involves a hierarchy of interdependent submodules, which requires multi-stage reasoning~\cite{tian2024scicode}. }


\hhy{Recent approaches have applied agent-based frameworks and tool-augmented coding systems, such as Claude Code CLI~\cite{claude}, to repository-level code generation. However, significant limitations remain. First, existing approaches are often inefficient because they involve redundant iterative reasoning and repeated context propagation across execution steps, leading to prohibitive computational costs. Our experimental results show that generating a thousand-line codebase with the state-of-the-art (SOTA) Claude Code CLI~\cite{claude} consumes more than ten million tokens.
Second, the lack of verification and error-recovery mechanisms substantially hinders reproducibility. \citet{poesia2022synchromesh} demonstrate that syntactic and type-related errors frequently arise when generating multi-file codebases. Similarly, \citet{hong2023metagpt} show that when models generate program logic and data structures simultaneously, redundant or inconsistent type definitions often emerge across files. These limitations highlight the need for systematic designs that support problem decomposition, state consistency, reasoning, and structured code generation.}


\hhy{To address these challenges, we propose \sys, an explicit multi-agent coordination system designed for reproducing network research code. We decompose the reproduction task into different stages, hierarchically deploying LLM-based agents to synthesize robust code artifacts.} Our system features a coordination mechanism that enforces a deterministic workflow, facilitating precise information exchange via a shared memory. Further, we implement a hierarchical repair strategy within isolated sandboxes: we first utilize static analysis to eliminate syntactic errors, followed by runtime debugging and semantic alignment to iteratively refine the generated code. \hhy{By integrating external bug-fixing tools with a LLM, \sys ensures executable code generation.}

\hhy{The key contributions of this paper are as follows:}
\begin{itemize}
    \item \textbf{A Multi-Agent Framework for Automated Code Generation}: \hhy{We propose \sys, a framework that orchestrates collaborative LLM agents to transform academic papers into executable code. By defining specialized roles---including \textit{Content Parsing Agent}, \textit{Architecture Design Agent}, \textit{Code Generation Agent}, and \textit{Audit \& Repair Agent}---we decompose complex generation tasks into manageable sub-processes, ensuring distinct responsibilities for extraction, design, implementation, and optimization.}
    \item \textbf{Efficient Context Management}: \hhy{To facilitate robust multi-agent collaboration, we design an explicit Shared Memory architecture that decouples information storage from the LLM reasoning process. We introduce a standardized data structure to manage both paper content and code hierarchies, coupled with a rule-based read--write protocol.} This mechanism ensures data consistency and enables seamless context sharing across different stages of the generation pipeline.
    \item \textbf{Sandbox Isolated Hierarchical Refinement Methodology}: We introduce an iterative refinement mechanism that integrates static code analysis with dynamic execution feedback. This framework enables agents to detect syntax errors via static checks and resolve logical runtime failures through execution traces, significantly improving the pass rate and reliability of the generated code.
    \item 
    We conduct extensive experiments to evaluate \sys. The results demonstrate that our approach significantly outperforms existing baselines in generating compile-ready and logically correct code from scientific literature. Moreover, with minimal human intervention, \sys achieves high-fidelity reproduction of the performance observed in official open-source implementations.
\end{itemize}

\section{Motivation}
\label{sec:motivation}
This section identifies the challenge of reproduction of network research results. \hhy{While the demand for reproducibility is high and LLMs are capable of code generation, current frameworks are too narrow in scope, limiting their applicability to network research.}

\subsection{Reproduction of Network Research is Imperative}

\begin{figure}[ht]
    \centering
    \includegraphics[width=0.95\linewidth]{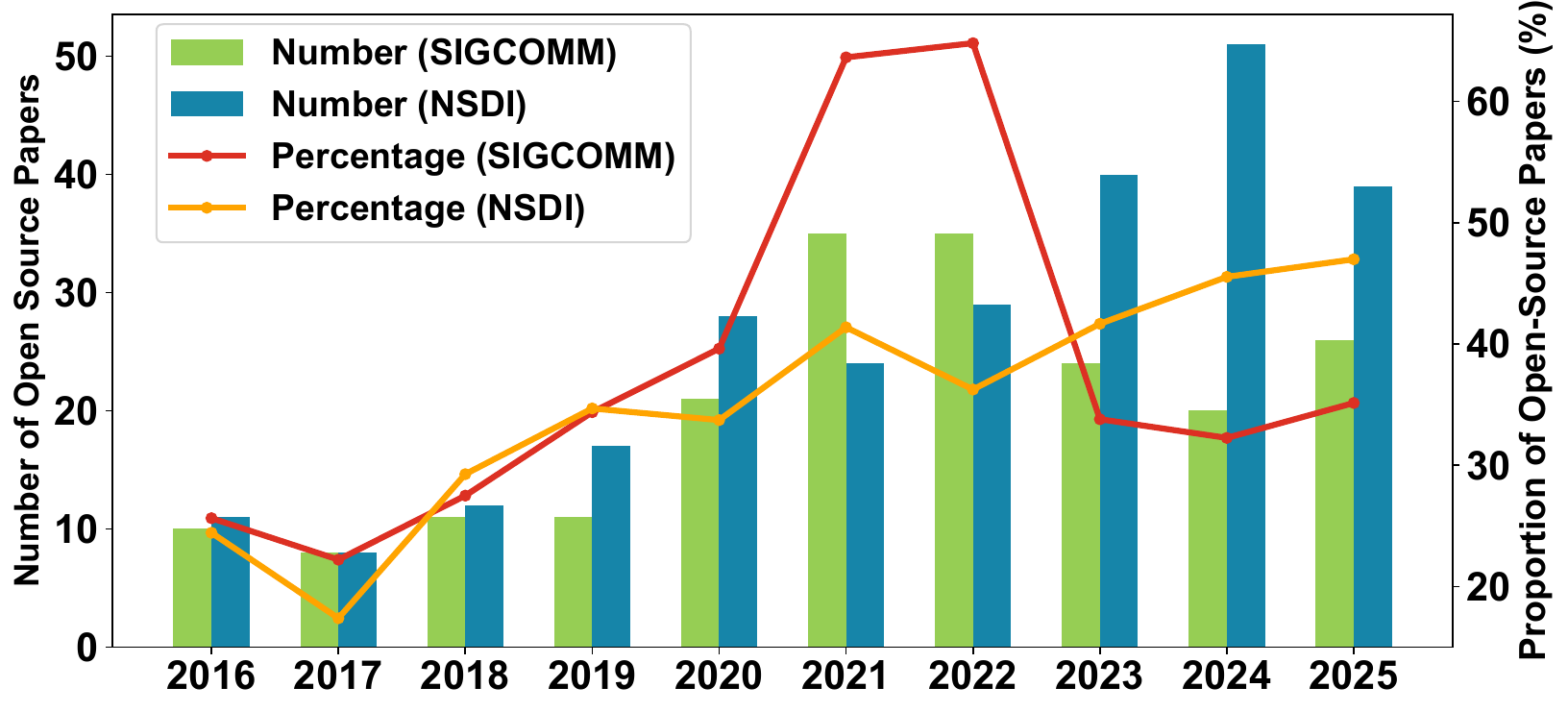}
    \vspace{-4mm}
    \caption{Statistics of SIGCOMM and NSDI papers with an
 open-source prototype from the authors (2016-2025).}
    \label{fig:open-source}
\end{figure}

\hhy{Conducting a study of network systems necessitates a rigorous comparison with SOTA solutions, requiring substantial effort from authors to reproduce prior work. As highlighted in~\cite{xiang2023toward}, 59.68\% of papers compare their contributions against at least two existing systems, with authors manually reproducing an average of 2.29 systems per publication.} Moreover, 49.20\% of these studies involved reproducing at least one comparison system, while 26.65\% required the reconstruction of two or more. 

\hhy{The manual reproduction of network system implementations presents several challenges. First, re-implementation imposes significant engineering overhead, diverting valuable time and resources away from novel research productivity. Second, the absence of source code or detailed experimental specifications can cause deviations from original results, undermining the credibility of comparative evaluations. Third, as modern network systems become increasingly complex, manual verification becomes less tractable, slowing the validation of new work and detracting from innovation.}


The limited availability of open-source implementations further exacerbates the overhead associated with reproducing experimental results. A statistical analysis of papers published in SIGCOMM and NSDI from 2016 to 2025, depicted in Figure~\ref{fig:open-source}, corroborates this observation: only 38.95\% of SIGCOMM and 37.37\% of NSDI papers made their source code publicly available during this period.

\subsection{Challenges of LLM-based Code Reproduction for Network Research}

\hhy{The heterogeneity of the computer networking field presents three primary challenges to automated research reproduction:}
\begin{itemize}
    \item \textbf{Diverse Research Topics:} \hhy{The field encompasses a wide array of distinct sub-domains (e.g., congestion control, traffic engineering, verification), challenging the domain-adaptability of LLMs.}
    \item \textbf{Unstructured Research Formats:} \hhy{Research papers lack a standardized format,  diverse paper organizations, and multimodal content like diagrams, challenging the interpretability of LLMs.}
    \item \textbf{Heterogeneous System Architecture:} \hhy{A network research work significantly varies from other proposed systems in system architecture or programming languages, thereby challenging the programming ability of LLMs.}
\end{itemize}

\hhy{The deficiency of existing methods for reproduction of network research motivates the development of our proposed new code generation frameworks. }

\subsection{Limitations of Existing LLM-based Code Generation Frameworks}
\hhy{Existing LLM-based code generation frameworks are designed either for general multi-task scenarios \cite{claude,hong2023metagpt,langchain,wu2024autogen} or domain-specific code generation frameworks \cite{seo2025paper2code, luo2025intention,li2025deepcode,zhao2025autoreproduce}.} We analyze the existing limitations in detail, which motivate our fine-grained LLM-based framework to automatic code generation of system proposed in network research.

\subsubsection{Limited Generalizability}
Existing code generation frameworks are limited to specific domains. \hhy{For example, \citet{seo2025paper2code, li2025deepcode} are for neural network (NN) models. \cite{luo2025intention} is designed for biomedical scenarios. These approaches use hard-coded domain-specific prompts for specific code generation. For example, in the NN model generation framework~\cite{li2025deepcode}, PyTorch interfaces, e.g., \texttt{nn.ReLU}, are strictly defined in prompts.}

\subsubsection{Excessive LLM Token Consumption}
\hhy{Existing general-purpose multi-task frameworks employ varying numbers of LLM agents to address diverse tasks. Due to the reliance on implicit coordination among agents, these frameworks suffer from excessive token consumption and ambiguous task allocation. }
This structural limitation manifests in two key dimensions:
\begin{itemize}
    \item \textbf{Inherent Redundancy in Multi-Agent Collaboration:} \hhy{Redundant loading and processing within the context window of every agent leads to significant resource inefficiency. 
    For example, AutoGen~\cite{wu2024autogen} utilizes broadcast mechanisms where identical messages are distributed to all participants.} 
    \item \textbf{Context Inflation in Long Tasks:} \hhy{Existing multi-agent systems, like Claude Code CLI~\cite{claude}, have exhaustive token consumption with naive context appending. Moreover, this accumulation dilutes relevant information and degrades model performance~\cite{wu2024autogen, claude_context_windows}.}
\end{itemize}



\subsubsection{Inefficient Content Extraction and Management}

\hhy{For content extraction, LLMs still struggle to efficiently interpret complex and structured academic documents. Conventional parsers often lose semantics when extracting formulas, algorithms, and other non-textual elements~\cite{blecher2023nougat}. Processing entire papers with long-context windows is also token-inefficient and prone to the ``Lost in the Middle'' effect, where models struggle to retrieve information embedded in dense contexts~\cite{liu2024lost}. Moreover, accurately segmenting documents while preserving links between figures, equations, algorithms, and their textual descriptions remains a key bottleneck~\cite{gao2023retrieval}.}


\hhy{For context management, existing code generation approaches often lack explicit structural schemas, causing LLMs susceptible to coherence degradation and context attrition over extended interactions. Previous studies have emphasized the role of context management in keeping code generation organized and consistent~\cite{jimenez2023swe}. Without structural constraints, LLMs may produce spurious file references, namespace collisions, circular dependencies, and redundant definitions. These problems collectively exacerbate hallucinations in LLM-based code generation~\cite{qian2024chatdev, hong2023metagpt, zhang2023repocoder}.}

\hhy{Moreover, LLMs remain susceptible to inconsistent or fragmented code generation because existing methods primarily depend on prompt-driven alignment~\cite{qian2024chatdev, hong2023metagpt, liu2025rtadev}. Prior studies show that expanding accessible context alone is insufficient to maintain long-range coherence~\cite{liu2024lost, ding2023crosscodeeval}. However, existing approaches often reduce context management to context-window expansion or retrieval optimization, without the structural consistency mechanisms required by systematic code generation.}

To address these limitations in context management, we propose content abstraction and external memory mechanism to keep the extracted data aligned with the generated code.

\subsubsection{Deficiency of Verification Mechanism}
\hhy{Generating code for large-scale network systems with LLMs in a ``one-pass'' manner is unreliable. Prior work~\cite{shinn2023reflexion} has shown that an iterative \emph{Generate $\to$ Check $\to$ Repair} loop is more effective for improving reliability. In automatic paper reproduction, generated code commonly exhibits two types of issues: \textbf{syntax-level issues}, where errors prevent the code from compiling or executing correctly; and \textbf{semantic-level issues}, where the implemented algorithm deviates from the original paper, data-processing workflows are incorrect, or input/output formats are mismatched. Such semantic issues may allow the code to execute successfully but still fail to produce the expected results.}

\hhy{Existing code generation frameworks often lack robust post-generation verification and debugging mechanisms. As shown in ~\cite{gulati2025controllable, yang2023intercode}, current models fail to detect incomplete or incorrect solutions, resulting in generation without effective iterative refinement. Although recent approaches have improved syntactic correctness, they often miss deeper semantic errors. Furthermore, LLM-generated code rarely contains syntax errors but still exhibits subtle logic flaws and functional deviations that conventional syntax checkers cannot detect.~\cite{wang2025characteristics, ni2023lever}}

\hhy{Moreover, existing verification mechanisms mainly target isolated, single-file unit tests and do not adequately capture repository-level dependencies. Although methods such as Self-Debug~\cite{chen2023teaching} leverage execution feedback, they cannot verify global consistency across interdependent modules. Without enforcement of architectural constraints, such as cross-module imports and symbol definitions, LLMs may generate code that is syntactically valid but structurally unbuildable~\cite{jimenez2023swe}.}

\hhy{These limitations motivate our runtime improvement framework. By combining static analysis with dynamic exception handling, our approach detects structural and runtime errors, refines implementation details, and improves the robustness of generated code.}

\section{System Overview}
\label{sec:overview}

\begin{figure}[ht]
    \centering
    \includegraphics[width=0.99\linewidth]{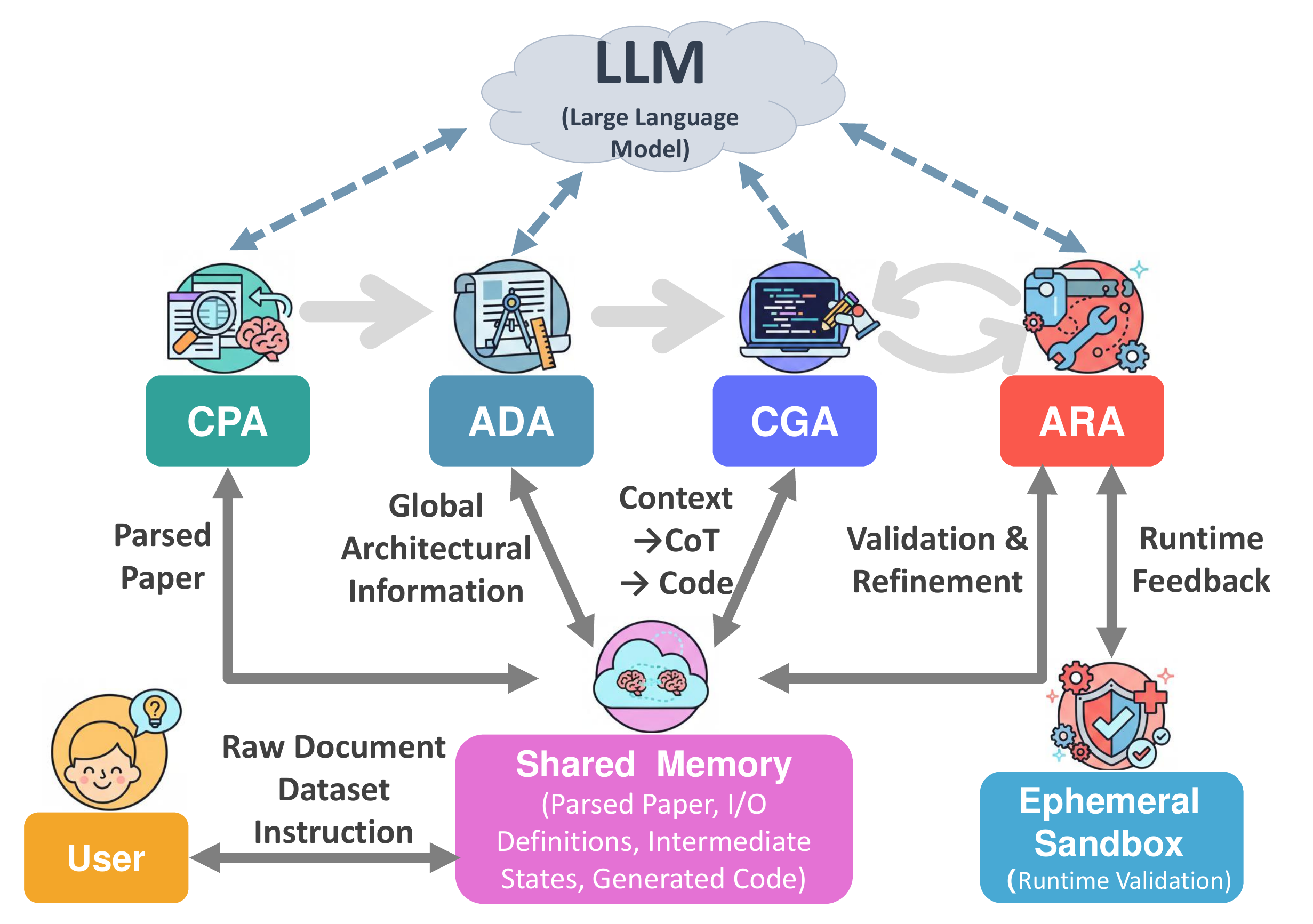}
    \vspace{-4mm}
    \caption{Overview of \sys's Multi-Agent Pipeline.}
    \label{fig:overview}
\end{figure}

We introduce \sys, a multi-agent framework designed for automated code generation of network system research. 
\sys decomposes the workflow into a pipeline: 1) \textbf{Content Parsing}, 2) \textbf{Architecture Design}, 3) \textbf{Code Generation}, and 4) \textbf{Code Audit \& Repair}. 
\hhy{We design four collaborated agents to achieve efficient code generation.} The overview of the pipeline code generation is shown in Figure~\ref{fig:overview}. \hhy{The pipeline proceeds as follows:}
\begin{itemize}
    \hhy{\item \textbf{Content Parsing Agent (CPA)} transforms raw research documents into structured information, explicitly managing multimodal information rather than narrative text.

    \item \textbf{Architecture Design Agent (ADA)} decomposes the target system into a directed acyclic graph (DAG) of operational steps, establishing explicit I/O interfaces between consecutive steps.
    
    \item \textbf{Code Generation Agent (CGA)} synthesizes executable code for each step, using structured intermediate representations to bridge the semantic gap between natural language descriptions and program logic.
    
    \item \textbf{Audit \& Repair Agent (ARA)} validates generated code through static analysis and runtime execution, feeding error feedback back into the generation loop for iterative refinement.}
\end{itemize}


\hhy{Moreover, we design a \textbf{Shared Memory} to manage extracted data from paper and the generated content of the agents. It coordinates the pipelined stages of code generation, thereby preventing semantic drift across the whole workflow.}

\subsection{Workflow of \sys} \label{sec:sequential_workflow}
\begin{figure}[ht]
    \centering
    \includegraphics[width=1.0\linewidth]{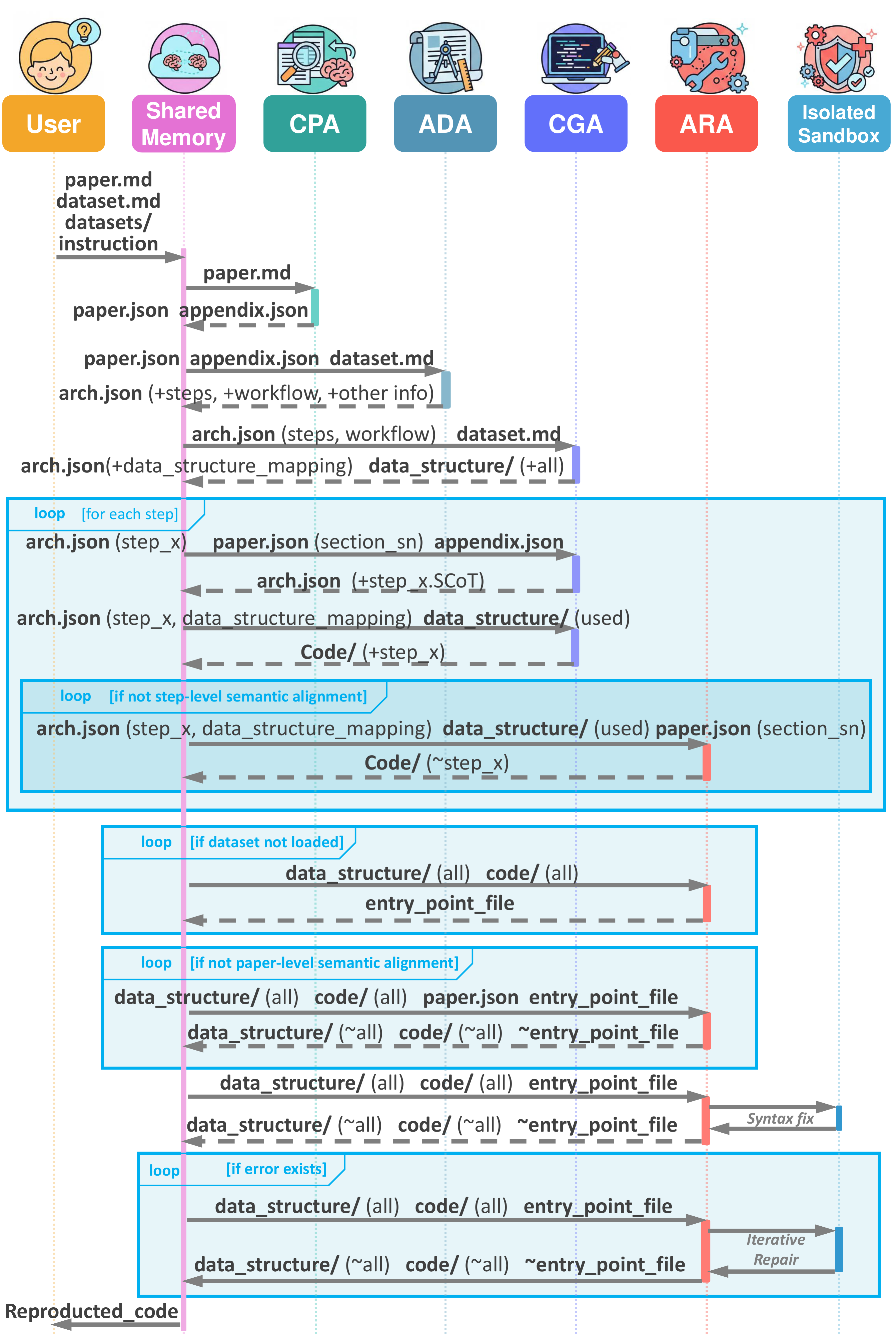}
    \caption{Workflow of RepLLM}
    \label{fig:workflow}
\end{figure}

The workflow of \sys is shown in Figure \ref{fig:workflow}, where $+$ denotes addition and $\sim$ denotes modification.

\subsubsection{Phase 1: Content Parsing \& Initialization}


The workflow begins with the User providing the raw research materials, including the manuscript (\texttt{paper.md}), dataset documentation (\texttt{dataset.md}), raw datasets (\texttt{datasets/}), and the user's instructions (\texttt{Instruction}). The \emph{CPA} ingests these documents. It extracts multimodal content (figures, tables, algorithms) and preserves mathematical equations in standard \LaTeX{} format. Then the \emph{CPA} serializes the content into a structured JSON representation (\texttt{paper.json} and \texttt{appendix.json}). This creates a standardized, machine-readable index of the paper's sections, cross-references, and multimedia elements. These structured files are stored in the \textit{Shared Memory}, serving as the ground truth for all subsequent agents.

\subsubsection{Phase 2: Architectural Design}


Once the content is parsed, \emph{ADA} accesses the \texttt{paper.json} and \texttt{dataset.md} from \textit{Shared Memory} to design the software architecture.\emph{ADA} decomposes the reproduction task into a Directed Acyclic Graph (DAG) of discrete steps. It generates a detailed description for each step, including input/output data structures and dependencies.\emph{ADA} outputs an \texttt{arch.json} file to \textit{Shared Memory}. This file contains the workflow steps, dependencies (\texttt{step\_x}), and mappings for data schemas (\texttt{data\_schema\_mapping}), and system information such as environment dependencies

\subsubsection{Phase 3: Code Generation Loop}

Before writing functional code, \emph{CGA} defines and generates the necessary class definitions and input/output interfaces, storing them in \texttt{data\_schema/}.

\vspace{0.5em}
The system enters a loop to generate code for each step defined in \texttt{arch.json}.
 For a specific \texttt{step\_x}, \emph{CGA} retrieves the relevant section context (\texttt{section\_sn}) and step definitions from \texttt{arch.json} and \texttt{paper.json}.
 \emph{CGA} first generates a Structured Chain-of-Thought (SCoT) (\texttt{step\_x.SCOT}). This intermediate representation maps out the logical flow before actual coding begins. Guided by the SCoT and data structures, \emph{CGA} writes the executable code for the step, saving it to the \texttt{code/} directory. 
 
The system immediately validates the generated code. If the code does not align semantically with the step description (detected via \emph{ARA}), a repair loop is triggered.
The agent reads the current code and context, fixes the discrepancies, and updates the files in \texttt{code/} and \texttt{data\_schema/}.

\subsubsection{Phase 4: Audit, Repair, \& Execution}

After all steps are generated, \emph{ARA} takes over to integrate, verify, and execute the code.
\emph{ARA} checks if the dataset loading logic is correct. It synthesizes a unified \texttt{entry\_point\_file} (e.g., \texttt{main.py} or bash script) to orchestrate the modular components. After that, \emph{ARA} performs a paper-level semantic check. It verifies that the integrated system logic matches the overall methodology described in the manuscript. If misaligned, it modifies the code and entry point file.
Before execution, \emph{ARA} runs static analysis tools (e.g., Ruff for Python) to detect and resolve syntax.
Then, based on \texttt{arch.json}, \emph{ARA} generates the Docker configuration, executes the code in a Docker container, and iteratively repairs errors identified during execution.

\section{Design of Agents}
\label{sec:design}
\hhy{This section presents the detailed design of the agents and the \emph{Shared Memory}.} 

\subsection{Content Parsing Agent}
\label{subsec:paper-parsing}
\hhy{\emph{CPA} leverages LLM to reorganize raw document content into a structured JSON representation. This approach standardizes interactions with the LLM, thereby enhancing the efficiency and accuracy of multimodal content comprehension. All extracted data is persisted within \emph{Shared Memory}.}

\subsubsection{Multimodal Content Extraction}
\hhy{We utilize the MinerU framework~\cite{wang2024mineru} to extract multimodal content, including figures, tables, algorithms, and equations, from the PDF file. Specifically, the framework preserves tables using standard Markdown format and references figures via generated image URLs. To ensure mathematical fidelity, both equations and algorithms are retained in their native \LaTeX{} representations. We transcode source document into Markdown formats, which facilitates JSON. By leveraging this tool, we are able to preserve information in a more comprehensive and faithful manner.}

\hhy{\subsubsection{Content Organization}
We index content globally for fast retrieval. We segregate auxiliary elements, i.e., figures, tables, algorithms, and equations, into distinct arrays, strictly preserving their original sequential order. Textual content is organized hierarchically: root-level entries correspond to top-level sections, with subsections nested 
within their respective parents. Bibliographic citations are explicitly excluded. 

Moreover, for multiple cross-reference results in a research paper, e.g., figures and algorithms among sections, \emph{CPA} explicitly maintains lists of referenced figures, tables, algorithms, and objects for extracted sections. 
When a downstream agent queries content for a specific section, \emph{CPA} recursively expands all cross-referenced content, ensuring that the agent receives a self-contained context without implicit external dependencies. This design avoids a common failure: code generation references an assumption mentioned three sections earlier but not included in the context window}






\subsection{Architecture Design Agent}
\label{subsec:architecture-design}
\hhy{\emph{ADA} orchestrates the structural decomposition and hierarchical organization of the generated codebase. 
\emph{ADA} maintains architectural consistency by explicitly recording dependencies and file mappings in a unified schema, which guides all subsequent code generation phases. 
Further, we incorporate an external \emph{Shared Memory} mechanism to manage LLM-generated content. 
It facilitates global code maintenance and ensures precise coordination across multiple files. Moreover, it significantly accelerates the generation process by reducing unnecessary token cost on redundant content.}

\subsubsection{Task Decomposition}
First, \emph{ADA} decomposes the comprehensive system reproduction task into discrete steps and utilizes distinct files to store step-wise code.
We model the logical dependencies among steps as a Directed Acyclic Graph (DAG). 

Within each step, \emph{ADA} generates step description details, including data structure specifications of input/output (I/O), and a section identifier of related extracted context. For each step, inputs and outputs explicitly indicate their data sources, whether from the original dataset or outputs of preceding steps.
These artifacts are serialized in JSON format and persisted in \emph{Shared Memory}. 
\emph{ADA} also leverages LLM to generate a high-level context summary, capturing the overall task, application domain, and expected I/O of target system from manuscript. 
Moreover, in order to constrain and facilitate downstream code generation, \emph{ADA} incorporates a manual specification of the target programming languages.

\subsubsection{Cross-File Dependency Management}
\hhy{\emph{ADA} organize content generated by LLMs through indices. 
We map LLM-generated artifacts, including source code files, I/O definitions, and their corresponding data schema definitions, with key-value pairs stored in \emph{Shared Memory}. 
In subsequent code generation phases, \emph{ADA} straightforwardly queries relevant context. 
This approach significantly reduces data retrieval overhead and preserves global coherence across the multi-file codebase.}

\subsection{Code Generation Agent}
\label{subsec:code-generation}
\hhy{\emph{CGA} generates code with two distinct phases: data schema definition and functional logic implementation. 
Before generating the core logic, \emph{CGA} synthesizes the necessary data schema code based on the I/O specifications from \emph{Shared Memory}. 
This preliminary step ensures interface consistency between steps and eliminates potential syntactic errors in advance. 
Further, \emph{CGA} utilizes Structured Chain-of-Thought (SCoT)~\cite{li2025structured} to systematically implement the functional code for each step.}

\hhy{\subsubsection{Code Generation Pipeline: Schema-first, logic-second}
\emph{CGA} first generates data schema definitions, such as \texttt{TypedDict} declarations for Python and \texttt{struct}/header definitions for C and P4, to precisely characterize the structure of data exchanged within each step. In this phase, field names, nested data structures, and type annotations are aligned with the dataset specification so that downstream data-loading components can parse actual inputs reliably and consistently.

Then, \emph{CGA} implements each step by treating the schema as an explicit contract and standardizing output using the corresponding schema types.
It is noteworthy that the strategy also addresses the I/O shifting problem in long-iteration LLM calling, where a data field's type or name diverges across steps.

\subsubsection{Structured Code Translation (SCoT)} 
We first employ SCoT to translate an algorithm into a structured notation, and subsequently translate this notation into target-language code. In this process, SCoT serves as an intermediate representation that bridges natural-language algorithm descriptions and executable code. Furthermore, for each code-generation step, we define operations over predefined data structures, where the input and output schemas explicitly specify the data dependencies required to coordinate the generation process.

SCoT captures the control flow of each step, such as sequential execution, conditional branching, and iterative looping, in a language-agnostic manner. By separating \textit{what to do} in SCoT from \textit{how to implement it} in code, SCoT reduces the reasoning burden on LLMs. In particular, the model no longer needs to concurrently interpret natural-language descriptions, determine program structure, and ensure data schema consistency during code generation.}

\subsection{Audit \& Repair Agent}
\hhy{\label{subsec:audit-repair}
\emph{ARA} achieves system orchestration and step-wise verification, leveraging a hybrid of static and dynamic repair strategies to ensure codebase consistency and correctness.

\subsubsection{System Orchestration}
\emph{ARA} first generates a global entry script for the entire system. The LLM is prompted to generate this script based on the complete system architecture, the generated step-level file tree, and the dataset structure.

The generated script includes four primary components: First, it specifies a well-defined execution pipeline for all modules according to the DAG produced by \emph{ADA}. Second, it includes commands for installing the necessary dependencies (e.g., pybatfish). Third, it provides command-line argument parsing to configure runtime parameters. Fourth, it performs pre-validation checks before execution. By combining these functions into a single entry script, \emph{ARA} mitigates common failure modes caused by multiple entry points, unclear inter-module dependencies, and mismatched runtime arguments.

\subsubsection{Static Repair} Static repair has three steps, i.e., step-wise verification, system-level semantic alignment, and static syntactic repair. 
\paragraph{Step-Wise Verification}
\emph{ARA} retrieves the generated code from \emph{Shared Memory} and cross-validates it against the corresponding text for each step-wise code generation. 
If discrepancies are detected, \emph{ARA} instructs LLM to update code with patches to enhance alignment with the specifications.
\paragraph{System-Level Semantic Alignment}
\emph{ARA} verifies the codebase's alignment with the corresponding text. 
It evaluates the semantic deviations between generated code (data schema definitions and step implementations) and the text description (paper content, appendix and dataset descriptions). 
\paragraph{Static Syntactic Repair}
To avoid unnecessary LLM invocations, \emph{ARA} first applies off-the-shelf static analysis tools before resorting to LLM-based repair. To accommodate different programming languages, \emph{ARA} provides language-specific analysis and repair interfaces. For example, for Python programs, we employ Ruff~\cite{ruff} to detect and automatically fix syntax issues.

\subsubsection{Dynamic Repair} \emph{ARA} utilizes an isolated sandbox to achieve dynamic repair, thereby enhancing robustness and safety. It has two steps, i.e., sandbox construction and iterative runtime repair.

\paragraph{Sandbox Construction}
\emph{ARA} leverages Docker~\cite{docker} to construct isolated sandbox environments for reproducible execution. Firstly, it generates Docker configuration files according to the system architecture and the generated codebase. For systems that depend on third-party environments, such as \texttt{p4-bmv2}, \emph{ARA} retrieves a pre-existing base image and configures the required runtime environment on top of it.

In addition, \emph{ARA} embeds the generated entry script into the Docker image and assigns the necessary executable permissions. The workspace, dataset, and output directories are mounted into the container, allowing the sandboxed environment to execute the entry script under controlled conditions.

\paragraph{Iterative Runtime Repair}
Iterative Runtime Repair is the final validation phase within \emph{ARA}, where the generated system is executed end-to-end to catch errors that are invisible to static analysis. Unlike syntax errors, runtime errors emerge only during execution: missing dependencies, environment misconfigurations, I/O interface mismatches, and unhandled edge cases. The repair operates inside the Docker sandbox and iterates up to five times.}

\hhy{\subsection{Shared Memory}
\label{subsec:shared_memory}

We utilize \emph{Shared memory} to manage information across all agents. It coordinates the pipelined code generation stages. We maintain a directory tree on disk. The shared memory holds three layers of information:

\begin{itemize}
    \item \textbf{Paper layer}: structured paper content extracted during ingestion, including cross-referenced sections, figures, tables, equations, and supplementary materials.
    \item \textbf{Architecture layer}: the system architectural blueprint produced by the design stage, which defines steps, their interfaces, the precedence graph governing execution order and the environment dependencies.
    \item \textbf{Code layer}: the generated implementation produced by the code synthesis stage.
\end{itemize}}

\section{Implementation}
\label{sec:implementation}

\hhy{This section presents the concrete mechanisms and algorithms that implement the design principles described in \cref{sec:design}. We focus on the data structures, parsing algorithms, patch protocol, and runtime pipeline that constitute the reproducibility engine.}

\subsection{LLM API Interaction}
\hhy{\label{sec:llm_api_interaction}
We adopt a stateless LLM API interface in which LLMs do not receive the full dialogue history in subsequent requests. 
Thus, there is no conversation history across interactions. All information provided to LLMs via API calls is explicitly manipulated by \sys, thereby achieving zero token waste.}

\subsection{Shared Memory}
\label{subsec:shared_memory}

\hhy{First, we implement \emph{Shared Memory} using the file system. Concretely, workspace contains a root-level \texttt{shared\_memory} directory that is accessible to all agents. In addition to the two structured artifacts described below, the workspace also includes all generated source files, such as module implementations, data schemas, configuration files, and intermediate outputs. Later stages can read and write these files under the same access conventions.

The dedicated \texttt{shared\_memory} directory is excluded from the file-tree construction during code generation. Thus, its intermediate representations are not exposed to LLMs as ordinary source files in prompts. This isolation improves both safety and robustness by preventing internal coordination artifacts from misinterpretation or modification as part of the generated codebase.

Second, we separate the \texttt{shared\_memory} into two file spaces: 1) the paper space for the content extracted from the paper and 2) the system space for the reproduced system architecture and code. 

For the paper space, we construct a \texttt{paper.json} file to store the extracted content in a structured tree. This tree organizes sections, figures, tables, algorithms, and equations, and is subsequently extracted into prompts for LLM-based generation.
Each section is assigned a unique identifier, which enables efficient access to relevant paper content. Correlated content is recursively linked through cross-references. In addition, an optional \texttt{appendix.json} file is attached to the section specified as the source of each generation step.

For the system space, we construct an \texttt{arch.json} file generated by \emph{ADA}. This file records the reproduced system architecture, including step decomposition, step-level workflows, interfaces, numeric contracts, and the environment dependencies. \texttt{arch.json} is updated in place during reproduction: data-schema generation appends schema mappings for each step, code generation writes per-step SCoT fields, and audit/repair generates the script for system entrance.

To handle incomplete intermediate states, missing files or referenced fields are replaced by all available results rather than treated as fatal errors. This design allows subsequent pipeline stages to remain executable while preserving a consistent system-level state.

Moreover, we implement \texttt{shared\_memory} as a coherent and agent-agnostic state layer. To avoid expired state, reads are in parallel with writes and are always resolved lazily from the latest on-disk content rather than from cached copies. When the resolved files need to be used, they are wrapped as fenced and language-tagged blocks and then embedded into LLM prompts. This design improves consistency across agents while reducing the risk of stale context being propagated through the generation pipeline.}
\subsection{SCoT}
\label{sec:impl-scot}
\begin{algorithm}[ht]
    \caption{SCoT Extraction from Paper Sections}
    \label{alg:scot-extraction}
    \begin{algorithmic}[1]
        \Function{Extract-SCoT}{step $s$, paper $P$, dataset description $D$}
            \State $c_{\text{section}} \gets \text{Recursive-Context-Retrieval}(P, s.\text{section\_sn})$
            \State $\text{ctx} \gets s.\text{description} \oplus c_{\text{section}} \oplus D$

            \Statex \qquad \textbf{// Stage 1: generate SCoT from natural language}
            \State $\text{SCoT}_{\text{raw}} \gets \text{LLM}(\text{ctx}, \text{SCoT\_prompt\_templates})$
            \State $\text{SCoT}_{\text{json}} \gets \text{Extract-JSON}(\text{SCoT}_{\text{raw}})$

            \Statex \qquad \textbf{// Stage 2: generate code from SCoT}
            \State $\text{schema} \gets \text{Get-Data-Schema}( s.\text{step\_index})$
            \State $\text{code}_{\text{raw}} \gets \text{LLM}(\text{SCoT}_{\text{json}} \oplus \text{schema})$
            \State $\text{code}_{\text{final}} \gets \text{Parse-JSON-Block}(\text{code}_{\text{raw}})$

            \State \Return $(\text{SCoT}_{\text{json}}, \text{code}_{\text{final}})$
        \EndFunction
    \end{algorithmic}
\end{algorithm}
\hhy{SCoT performs a two-stage translation: first from section text to structured SCoT, then from SCoT to executable code. Applied in \emph{CGA}, Algorithm~\ref{alg:scot-extraction} illustrates the two‑stage process. Given a step information, the paper, and the dataset description, it first assembles the context by retrieving the relevant section content. Stage 1 prompts the LLM to generate a raw SCoT, then parses it into a JSON structure and writes it into \textit{arch.json}. Stage 2 feeds the SCoT together with the required data schema back to the LLM to produce raw code, which is finally parsed into executable code. The algorithm returns the SCoT along with the final generated code.

SCoT explicitly provides formal interface contracts (i.e., typed input/output specifications) and orchestrates the underlying logic using language-agnostic control flow notations, such as sequential steps, explicit branching, and loop blocks. 

Moreover, SCoT abstracts algorithmic behavior through highly structured natural language, rather than mimicking programming syntax with mathematical notation, language-specific API conventions, or implicit data structures as in traditional pseudo-code.
From an architectural perspective, SCoT provides an intermediate representation that bridges natural-language algorithm descriptions and concrete source code.}

Figure~\ref{fig:scot_topo_parser} illustrates an example of SCoT for Mininet topology parsing. It first specifies the typed input and output of the step, where \texttt{topo\_file} denotes the path to the topology description and \texttt{topo\_obj} denotes the constructed \texttt{Topo} object. Then, it describes the logic as an ordered sequence of structured operations, including importing the required topology abstraction, defining a custom topology class, reading the topology file, parsing each line into node pairs, and adding the corresponding links. The loop and conditional branches explicitly capture the control flow, while avoiding commitment to low-level implementation syntax.

Moreover, line~4 uses the language-agnostic expression ``Open \texttt{topo\_file} for reading,'' which does not constrain the second-stage code generator to a specific implementation, such as Python's \texttt{with open} or C++'s \texttt{ifstream}. In this way, SCoT keeps the fundamental algorithm and data dependencies while leaving language-specific realization to the code generation stage.
\begin{figure}[ht]
    \centering
    \footnotesize
    \begin{lstlisting}[mathescape=true, escapeinside={(*}{*)}]
Input: topo_file: string representing the path to topo.txt
Output: topo_obj: an instance of Mininet Topo class
1:  Import Topo from mininet.topo
2:  Define a class CustomTopo that inherits from Topo
3:  Define the build method in CustomTopo that takes topo_file as an argument
4:  Open topo_file for reading
5:  for each line in the file do
6:      Strip whitespace and split the line into a list of tokens
7:      if tokens is empty then
8:          Continue to the next line
9:      else
10:         Call addLink(tokens[0], tokens[1])
11: Create an instance of CustomTopo with topo_file and assign it to topo_obj
12: return topo_obj
\end{lstlisting}
    \vspace{-6mm}
    \caption{An Example of SCoT for Mininet Topology Parsing}
    \label{fig:scot_topo_parser}
\end{figure}

\hhy{\subsection{Structured Data Extraction from LLM Responses}
\label{sec:impl-xml-extraction}
\sys reliably extracts machine-readable structured data and code from free-form text of LLM responses. 
We define a lightweight XML-like tagging protocol using three tag types: \texttt{<json>} for structured data, \texttt{<file>} for full source file content, and \texttt{<patch>} for targeted text replacements. All tags are case-insensitive, accept optional attributes in flexible order, and support HTML-escaped variants (\texttt{\&lt;json\&gt;}).

To robustly parse LLM outputs, \sys{} avoids brittle regex matching and instead uses a character-level scanner that recognizes closing tags only outside quoted strings. JSON extraction is performed through progressive fallbacks, including tagged blocks, HTML-unescaped tags, fenced code blocks, and bare JSON streams, with automatic repair for common LLM formatting errors. File and patch blocks are normalized before use, and patches are applied only when the target text is uniquely matched. Moreover, multiple edits to the same file are committed transactionally, with SHA-256 deduplication preventing duplicate applications.}

\section{Evaluation}
\label{sec:evaluation}
\xyx{In this section, we compare our proposed framework with SOTA code generation approaches. }To demonstrate the practical utility and robustness of our method, we present experimental results centered on the reproduction of network research papers published in top-tier academic conferences. 

\subsection{Baselines}
\xyx{To demonstrate the effectiveness of the \sys framework, we conduct a comparative evaluation against two representative baselines with distinct characteristics:}

\noindent\textbf{Standalone LLM}: We directly feed the entire raw documents to a LLM and prompt it to generate the complete codebase in a single pass.

\noindent\textbf{Claude Code CLI}~\cite{claude}: \xyx{Claude Code CLI is a command-line tool that provides AI-powered coding assistance, supporting autonomous file operations and bash execution. We employ Claude Code CLI to achieve multi-agent-based code generation.}

\xyx{We deploy Claude Code CLI in a command-line environment and configure multiple agents for code generation, debugging, and refactoring, respectively. To ensure full automation, we strictly constrain human involvement during the generation process: human operators only approve execution and file-modification requests, without providing additional guidance, hints, or manual intervention. We evaluate Claude Code CLI under three configuration modes, namely low, medium, and high, to assess its code generation performance.}

\subsection{Setup}

\xyx{For the LLM API, we use \texttt{Gemini-3.1-Pro-preview} as the backbone model. We employ Docker to conduct sandboxes. All experiments are conducted in a controlled environment equipped with a standard development stack and the necessary runtime dependencies. To improve evaluation stability and reduce run-to-run variance, We repeat each experiment on \sys{} three times and report the aggregated results.}

\subsubsection{Selected Research Paper}
\xyx{Our framework can generalize to any system-level paper with sufficient implementation detail, regardless of venue. Here we select seven representative papers spanning diverse networking research domains and architectures. As shown in Table~\ref{tab:selected_paper}, these works cover traffic engineering, DNS verification, coding theory, configuration mining, probabilistic verification, and in-network systems, and span implementations originally written in Python, C++, Go, and Python\&P4. To eliminate language-induced variance, we set the implementation of all frameworks and \sys exclusively to Python and P4.}
\begin{table}[!t]
    \centering
    \caption{\xyx{Selected Papers}}
    \vspace{-4mm}
    \label{tab:selected_paper}
    \footnotesize
    \resizebox{\columnwidth}{!}{%
    \begin{tabular}{c|c|c|c|c}
    \toprule
    \textbf{System} & \textbf{Conference} & \textbf{Area} & \textbf{Lang.} & \textbf{Tools} \\
    \hline
    NCFlow~\cite{abuzaid2021contracting} & NSDI'21 & Traffic Eng. & Python & Gurobi \\
    GRooT~\cite{kakarla2020otot} & SIGCOMM'20 & DNS Verif. & C++ & -- \\
    Rateless IBLT~\cite{yang2024practical} & SIGCOMM'24 & Coding Theory & Go & -- \\
    SelfStarter~\cite{kakarla2020finding} & NSDI'20 & Config Mining & Python & PyBatfish \\
    NetDice~\cite{steffen2020probabilistic} & SIGCOMM'20 & Prob. Verif. & Python & -- \\
    Teal~\cite{xu2023teal} & SIGCOMM'23 & Traffic Eng. & Python & CUDA \\
    NetChain~\cite{jin2018netchain} & NSDI'18 & In-network Sys. & Python\&P4 & Mininet \\
    \bottomrule
    \end{tabular}%
    }
\end{table}

\subsubsection{Evaluation Metrics}\label{subsec:metrics}
\xyx{We define a multi-dimensional set of metrics to evaluate the code generation performance of all methods:}
\begin{itemize}
    \item \textbf{Code-Level Reliability}: This metric assesses the syntactic correctness and execution readiness of the generated code. It consists of two sub-metrics: \xyx{1) \textit{Executability} and 2) 
    \textit{Code-Level Data Loading Success}, which verifies whether the generated code is able to load the open-source datasets referenced in the original paper.}

    \xyx{\item \textbf{System-Level Paper--Code Semantic Alignment}: This metric evaluates the structural and logical fidelity of the generated system with respect to the paper description. We adopt an LLM-based evaluation protocol to assess three aspects: completeness (Comp.), correctness (Corr.), and maintainability (Maint.).}

    \xyx{\item \textbf{Code Generation Cost}: This metric quantifies the computational and monetary cost incurred during code generation. We report the input tokens, output tokens, and cumulative token consumption across the entire generation process.}

\end{itemize}

\xyx{We further evaluate whether the generated code can reproduce the reported results. In this evaluation, we allow human intervention to correct semantic mismatches between the generated code and the target functionality.}

\begin{itemize}
    \item \textbf{Reproduction Performance}:\xyx{ We measure the consistency between the experimental results produced by the generated code and two reference sources: 1) the results reported in the original paper, and 2) the results obtained by executing the official open-source implementation.

    \item \textbf{Reproduction Calibration Cost}: We quantify the human effort required to refine the generated code into a valid reproduction. We measure the total refinement time (time-to-completion) and the number of modified lines (diff size), using Cursor IDE~\cite{cursor} for code editing.}
\end{itemize}

\subsection{Results of Code-Level Reliability}

\xyx{As shown in Table~\ref{tab:eval}, \sys achieves both executable and successful dataset loading on all tasks. In contrast, the Standalone LLM baseline fails on several tasks, including Rateless IBLT, Teal, and NetChain, where it cannot execute the system or properly integrate datasets, highlighting its limitations in robust environment reconstruction.

Moreover, Claude Code CLI also executes all tasks without runtime errors under three modes; however, it does not faithfully reconstruct dataset loading configurations. It usually provides over-simplified implementations or implicit assumptions of data rather than accurate integration, thereby limiting the availability to real datasets. For example, in NetChain,  Claude Code CLI produces a pure Python state-machine simulation, which entirely bypasses the BMv2 and Mininet.}
\begin{table}[!t]
    \centering
    \caption{\xyx{Evaluation of Code Executability and Dataset Loading}}
    \small
    \vspace{-4mm}
    \label{tab:eval}
    \resizebox{0.48\textwidth}{!}{
    \begin{tabular}{c|c|c|c|c|c|c}
    \toprule
        ~ & \multicolumn{2}{c|}{\textbf{Standalone LLM}} 
          & \multicolumn{2}{c|}{\textbf{Claude Code CLI}} 
          & \multicolumn{2}{c}{\textbf{\sys (ours)}} \\ 
        \cline{2-3} \cline{4-5} \cline{6-7}
        ~ & Runnable & \makecell{Dataset\\Loaded}
        & Runnable & \makecell{Dataset\\Loaded}
        & Runnable & \makecell{Dataset\\Loaded} \\ \hline
        Rateless IBLT & \xmark & \xmark  & \cmark & \cmark & \cmark & \cmark \\ 
        GRooT & \cmark & \cmark & \cmark & \cmark & \cmark & \cmark \\ 
        NCFlow & \cmark & \cmark & \cmark & \cmark & \cmark & \cmark \\ 
        SelfStarter & \cmark & \cmark & \cmark & \cmark & \cmark & \cmark \\ 
        NetDice & \cmark & \xmark & \cmark & \cmark & \cmark & \cmark \\
        Teal & \xmark & \xmark & \cmark & \cmark & \cmark & \cmark \\
        NetChain & \xmark & \xmark & \cmark & \cmark & \cmark & \cmark \\
    \bottomrule
    \end{tabular}
    }
\end{table}

\subsection{Results of Paper--Code Semantic Alignment}
Table~\ref{tab:judge_scores_avg} \xyx{summarizes the average paper--code semantic alignment scores across the seven benchmark papers. Overall scores are judge-provided holistic ratings and are not computed from the preceding three dimensions. The final column reports the equal-weighted average of the three judge-provided Overall scores.
The three LLM judges exhibit different calibration scales: GPT-5.5 assigns more conservative absolute scores, whereas Gemini-3.1-Pro-Preview tends to give higher scores. }

\xyx{Moreover, we report the equal-weighted cross-judge overall score and further characterize inter-rater reliability. 
We use Kendall's $W$ to measure rank agreement and ICC(3,1) with ICC(3,k) to measure consistency for a single judge and the averaged multi-judge score, respectively.

As shown in Table~\ref{tab:judge_agreement}, the three judges show consistent evaluation behavior despite their different score scales. 
For the holistic overall score, Kendall's $W$ reaches 0.747, indicating substantial rank agreement among the judges. 
The consistency-based ICC values further support this conclusion: ICC(3,1) is 0.593 for a single judge, while ICC(3,k) increases to 0.814 when aggregating the three judges. 
These results suggest that the relative quality assessment is stable across heterogeneous LLM evaluators, and the final averaged score is not dominated by the bias of a single judge.

\sys demonstrates strong reliability across repeated runs. 
Across three independent generations, \sys obtains a cross-judge averaged overall score of $6.18 \pm 0.24$, showing low run-to-run variance. 
This stability is important because paper-to-code generation is inherently open-ended: a reliable system should not depend on a single favorable generation. 
The small standard deviation indicates that \sys repeatedly produces repositories with comparable semantic alignment quality.

\sys also consistently improves over standalone LLM generation. 
Its averaged Overall score is $6.18$, compared with $5.40$ for Standalone LLM, corresponding to a $0.78$ point improvement. 
This gain, together with the low variance across repeated runs, indicates that \sys provides a more dependable generation process than directly prompting an LLM to generate the repository in one pass.

Overall, the results show that \sys's advantage lies not only in its score level but also in its robustness. 
The multi-judge agreement analysis supports the reliability of the LLM-based evaluation, while the repeated-run results show that \sys can stably produce semantically aligned code across independent executions.}

\begin{table*}[!ht]
\centering
\caption{\xyx{Average LLM-as-a-judge scores across seven papers.}}
\vspace{-4mm}
\scriptsize
\setlength{\tabcolsep}{3pt}
\resizebox{\textwidth}{!}{
\begin{tabular}{lrrrrrrrrrrrrr}
\toprule
Method
& \multicolumn{4}{c}{GPT 5.5}
& \multicolumn{4}{c}{Gemini-3.1-Pro-Preview}
& \multicolumn{4}{c}{DeepSeek-V4-Pro}
& \multicolumn{1}{c}{Cross-Judge} \\
\cmidrule(lr){2-5}\cmidrule(lr){6-9}\cmidrule(lr){10-13}\cmidrule(lr){14-14}
& Corr. & Comp. & Maint. & Overall
& Corr. & Comp. & Maint. & Overall
& Corr. & Comp. & Maint. & Overall
& Avg. Overall\\
\midrule
Standalone LLM 
& 3.00 & 5.29 & 4.57 & 4.04
& 7.00 & 7.86 & 6.71 & 7.26
& 4.29 & 5.86 & 5.00 & 4.89
& 5.40 \\

Claude Code CLI(low) 
& 3.00 & 5.57 & 5.29 & 4.27
& 7.86 & 9.29 & 8.29 & 8.07
& 5.57 & 7.14 & 6.14 & 5.80
& 6.05 \\

Claude Code CLI(medium) 
& 3.00 & 5.43 & 5.43 & 4.46
& 7.43 & 9.00 & 8.29 & 7.90
& 6.57 & 8.29 & 7.29 & 6.67
& 6.34 \\

Claude Code CLI(high) 
& 3.14 & 5.57 & 5.29 & 4.47
& 7.86 & 9.00 & 8.86 & 8.29
& 5.57 & 7.57 & 6.57 & 6.00
& 6.25 \\

Source Code 
& 5.71 & 7.00 & 6.57 & 6.44
& 9.43 & 9.43 & 9.43 & 9.31
& 8.86 & 9.29 & 8.43 & 8.96
& 8.24 \\

\sys (avg. $\pm$ std.) 
& 3.10$\pm$0.22 & 5.33$\pm$0.30 & 4.81$\pm$0.08 & 4.31$\pm$0.19
& 8.10$\pm$0.36 & 8.86$\pm$0.14 & 7.52$\pm$0.87 & 8.16$\pm$0.47
& 5.52$\pm$0.54 & 7.10$\pm$0.30 & 5.81$\pm$0.33 & 6.06$\pm$0.52
& 6.18$\pm$0.24 \\
\bottomrule
\end{tabular}
}
\label{tab:judge_scores_avg}
\end{table*}

\begin{table}[!ht]
\centering
\caption{\xyx{Inter-rater reliability among GPT-5.5, Gemini-3.1-Pro-Preview, and DeepSeek-V4-Pro judges.}}
\small
\begin{tabular}{lrrrr}
\toprule
Metric & Corr. & Comp. & Maint. & Overall \\
\midrule
Kendall's $W$ & 0.666 & 0.595 & 0.649 & 0.747 \\
ICC(3,1) & 0.559 & 0.528 & 0.517 & 0.593 \\
ICC(3,k) & 0.791 & 0.771 & 0.763 & 0.814 \\
\bottomrule
\end{tabular}
\label{tab:judge_agreement}
\end{table}

\subsection{Results of Code Generation Token Cost}
\begin{figure}[t]
    \centering
    \includegraphics[width=1.0\linewidth]{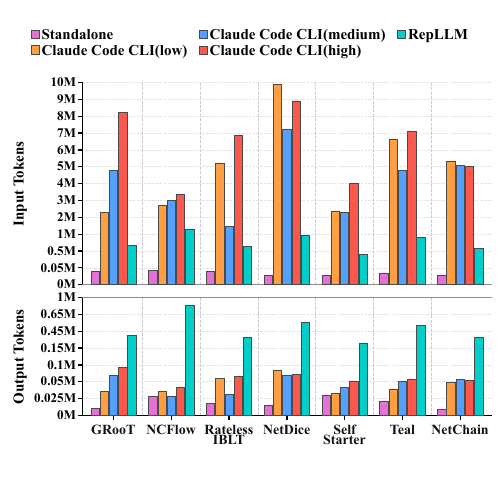}
    \vspace{-10mm}
    \caption{\hhy{Code Generation Token Cost of Different Frameworks.}}
    \label{fig:token_cost}
\end{figure}

\xyx{Figure~\ref{fig:token_cost} reports token consumption across baselines. 
For \sys, we report mean $\pm$ standard deviation to show run-to-run consistency. From the results, baseline methods either generate little code or require very large input budgets. 
In contrast, \sys achieves a better balance between input cost and effective code generation. The result demonstrates that \sys is more efficient in system-level code generation with a given budget.

Standalone LLM consumes few tokens, but they also produce limited code. For example, on GRooT, the standalone LLM uses only 38.4K input tokens and generates 10.3K output tokens. This low cost mainly reflects limited system coverage rather than high efficiency. 

Claude Code CLI incurs substantially higher input costs across benchmarks. On NetDice, Claude Code CLI requires roughly 9M input tokens, respectively. These large input budgets reflect repeated context processing across different modes, yet the final output remains much smaller than that of \sys. 

In contrast, \sys keeps input consumption substantially below Claude Code CLI while producing much larger implementation-oriented outputs. For instance, on NetDice, it generates 558.4K output tokens with 958.1K input tokens. These outputs include not only executable code, but also structured intermediate artifacts such as pseudo-code, architecture design, patches for improvement, and sandbox execution scripts.}

\subsection{Reproduction Calibration Cost}
\begin{table}[!t]
    \centering
    \setlength{\tabcolsep}{2.2pt}
    \caption{\xyx{Cost Analysis of Code Generation and Repair}}
    \label{tab:cost_t}
    \small
    \begin{tabular}{c|c|c|c|c}
    \toprule
        ~ 
        & \shortstack{\textbf{Generation}\\\textbf{Time}\\\textbf{(min)}} 
        & \shortstack{\textbf{Human Fixing}\\\textbf{Time}\\\textbf{(min)}} 
        & \shortstack{\textbf{Total}\\\textbf{Time}\\\textbf{(min)}} 
        & \shortstack{\textbf{Modified}\\\textbf{Lines}} \\
    \hline
        \textbf{GRooT} & 50  & 76 & 126 & 1535 \\ 
        \textbf{NCFlow} & 42 & 93 & 135 & 53 \\ 
        \textbf{Rateless IBLT} & 38 & 84 & 122 & 428 \\ 
        \textbf{NetDice} & 63 & 66 & 129 & 764 \\ 
        \textbf{SelfStarter} & 41 & 41 & 82 & 208 \\ 
        \textbf{Teal} & 59 & 23 & 82 & 232 \\ 
        \textbf{NetChain} & 84 & 54 & 138 & 1174 \\ 
    \bottomrule
    \end{tabular}
\end{table}
 
\xyx{Table~\ref{tab:cost_t} reports the reproduction cost of \sys{}, measured by code generation time, human repair time, total time, and modified lines. The automated code generation stage takes 38--84 minutes. The remaining human repair takes 23--93 minutes. Across all benchmarks, end-to-end reproduction of results finishes within 82--138 minutes. Most of the later human modifications were needed because the paper describes the protocol and algorithms, but does not provide full artifact-level implementation details. For example, in Netchain, performance-analysis scripts had to be added manually because the paper reports evaluation results but does not specify the exact scripts. Only a few changes were due to semantic mismatches, such as the exact NetChain wire-field order: the paper gives only a schematic packet-format diagram, not a byte-level implementation specification.
The results show that \sys{} can quickly provide a strong implementation baseline and make full-system paper reproduction practical with modest human effort.}

\xyx{\subsection{Analysis of Reproduced and Official Implemented Performance}\label{subsec:results_of_performance}
We analyze the reproduction performance of our proposed \sys against official implementations. Due to the page limit, the detailed analysis and results for Claude Code CLI and standalone LLM are in Appendix~\Cref{sec:appd_reproduction_analysis}.

\subsubsection{Algorithm-only systems: GRooT, Rateless IBLT, and NetDice}
Compared with standalone LLM baselines and Claude Code CLI, \sys{} better recovers paper-specific system structure without any function-level approximation. Although few semantic inconsistencies remain, after manual auditing and correction, \sys{} achieves substantially ``perfect'' reproductions: GRooT matches the benchmark violation profile on reference tests, Rateless IBLT validates the real protocol end to end, and NetDice passes all tests after correcting hot-edge reasoning, ECMP semantics, and bounded-exploration logic.

\subsubsection{Third-party-tool-dependent systems: NCFlow and SelfStarter}

Compared with the baselines, \sys{} more accurately reconstructs the required system decomposition. The remaining errors mainly arise from external-tool interfaces and ambiguous data formats. This is caused by the gap between the paper-level description and the implementation-level data interface. 
After manual correction, NCFlow achieves near-identical total flow on the main dataset, and SelfStarter matches the expected ACL-level profile, including template counts, group outliers, parameter-level discrepancies, and final consistency summaries.

\subsubsection{Hardware-dependent system: Teal}

Compared with the standalone LLM and Claude Code CLI baselines, \sys{} better preserves the official system design, whereas the baselines either fail during data loading or reduce the learning task to simplified surrogate experiments. The remaining limitations mainly stem from framework and hardware constraints. After manual correction, \sys{} faithfully executes the FlowGNN--COMA--ADMM pipeline on B4, completes UsCarrier when path caches are available, produces valid split ratios, and reports the expected traffic-engineering metrics, including measurable ADMM gains over policy-only allocation.

\subsubsection{Distributed runtime system: NetChain}
NetChain requires a P4 data plane, a Python control plane, clients, and orchestration scripts to run jointly on Mininet/BMv2. This setting stresses distributed component closure: all modules must be present, share consistent topology and protocol state, and execute within the intended runtime environment.

\sys{} is the only approach that approaches true end-to-end NetChain reconstruction. Although \sys{} initially generates a four-module framework with imperfect client and protocol logic, the human-corrected version fixes lifecycle management and resource cleanup, and successfully runs packet exchange on the BMv2 testbed with consistent client/receiver throughput logs.}

\xyx{\subsection{Ablation Studies}
We conduct ablation studies to evaluate the effectiveness of our three core components: SCoT, Shared Memory, and \emph{ARA} on reproducing SELFSTARTER.}
\begin{table}[H]
    \centering
    \setlength{\tabcolsep}{1.5pt}
    \caption{\xyx{Ablation Study Results}}
    \label{tab:ablation_results}
    \footnotesize

    \begin{tabular}{c c c}
    \toprule
        \textbf{Component} 
        & \textbf{Runnable} 
        & \textbf{Dataset Loaded} \\
    \midrule

        w/o SCoT & \cmark & \xmark \\ 
        Pseudocode & \xmark & \xmark \\ 
        w/o Shared Memory & \xmark & \xmark \\ 
        w/o Architecture Memory & \cmark & \xmark \\ 
        w/o Structured Paper Memory & \cmark & \cmark \\ 
        w/o Interface Memory & \xmark & \xmark \\ 
        w/o ARA (Static) & \cmark & \cmark \\ 
        w/o ARA (Dynamic) & \xmark & \cmark \\ 

    \bottomrule
    \end{tabular}
\end{table}

\subsubsection{Impact of SCoT}

\xyx{We compare the full pipeline with two ablations: \textit{w/o SCoT}, which removes SCoT, and \textit{Pseudocode}, which replaces it with conventional pseudo-code interleaved with code. Both variants discard data schemas, interface contracts, and per-step algorithm traces. Consequently, neither reproduces SELFSTARTER end to end, despite consuming 828K and 386K tokens, respectively, exceeding the full pipeline.

The failures are consistent: repair patches inject prose into source files, core algorithms are omitted after repeated repair failures, and the configuration parser binds to an unreachable external service instead of using lightweight fallbacks. These results show that SCoT is a structured contract between high-level plans and executable artifacts, providing the audit-repair loop with actionable signals that lightweight substitutes fail to preserve.}




\xyx{\subsubsection{Impact of Shared Memory}
We evaluate four shared-memory ablations, each removing a different component of cross-stage information flow. 
On SELFSTARTER, all variants exhibit the same failure pattern: apparent execution success with semantically incomplete outputs. Without Shared Memory, the paper context is reduced by two orders of magnitude, yielding a trivial script that succeeds by only checking for an output directory. Without Architecture Memory, the system produces a runnable scaffold but leaves core modules unimplemented. Without Structured Paper Memory, the parser relies on mocked behavior and generates hollow summaries despite consuming 1.1M tokens. Without Interface Memory, later modules drift from earlier interface commitments, causing failed configuration parsing to be treated as success. These results show that token efficiency alone is misleading in long-horizon agentic reproduction; the memory layer is essential for enforcing semantic completeness and cross-stage consistency under context constraints.


\subsubsection{Impact of ARA} 
We ablate the two repair loops in the \emph{ARA}: the static pass, which audits generated code for semantic consistency, and the dynamic pass, which executes the startup script in a sandbox and repairs runtime failures. On SELFSTARTER, the two passes address complementary errors. Without Static Repair, the code remains executable but suffers from semantic inaccuracies, such as imprecise ACL line matching and missed detections in no-clear-majority cases where devices report conflicting parameter values. Without Dynamic Repair, semantically plausible code fails at deployment due to sandbox-only issues, including Batfish port conflicts, missing container images, and orchestration failures. These results show that semantic correctness and executable deployment require distinct verification mechanisms, and that \emph{ARA}'s two repair loops jointly convert borderline implementations into faithful and runnable reproductions.}


\section{disscussion}
\xyx{\subsection{Limitations}
\label{sec:discussion}
\paragraph{Fragility of TypedDict.}
TypedDict keeps data as plain dicts with no implicit behaviour binding, maintaining pipeline flexibility at the cost of having no runtime enforcement: schema violations --- dropped \texttt{None}s, renamed keys, mistyped fields --- propagate until a runtime assert catches them, and the catch rate is bounded by the iteration cap, not by validation correctness.

\paragraph{Fragility of regex- and XML-based LLM response parsing.}

\sys{} uses a lightweight tag-based parser to extract JSON architectures, file blocks, and code patches from raw LLM responses. Even though this parser can handle common formatting deviations, including HTML-escaped tags and fenced-code fallbacks, its flexibility also makes the pipeline fragile. Stray closing tags, missing attributes, or explanatory prose may cause valid edits to be dropped or matched ambiguously. Constrained decoding or grammar-validated schemas could improve robustness.


\paragraph{Iteration caps and their justification.}
\sys{} bounds each repair loop with small fixed limits: the static-audit loop and the runtime-repair loop run at most three and five iterations, respectively. These limits are practical safeguards against non-progressing repair cycles. For example, malformed code patches may cause the patcher to repeatedly reject edits, preventing progress on the underlying bug. By failing fast and skipping stuck steps, \sys{} keeps the overall pipeline budget and allows downstream stages and final evaluation to proceed. The trade-off is that some recoverable bugs may be abandoned too early; we therefore prioritize system availability over exhaustive repair.


\paragraph{Scope of reproducible paper.}
In practice, \sys{} has limited support for three classes of systems papers: 1) \textit{Kernel- and driver-level systems.} Papers that require patched kernels, custom eBPF programs, or GPU driver extensions are difficult to compile and test in the lightweight Docker sandbox provisioned by \emph{ARA}. Their long build times, complex toolchains, and privileged execution requirements often prevent the audit loop from producing useful feedback. These failures are primarily environmental rather than algorithmic. 2) \textit{Hardware-dependent and multi-node systems.} Papers that rely on specialized hardware or large-scale distributed deployments, such as custom NICs, accelerators, RDMA clusters, multi-rack BGP topologies, or geo-distributed key-value stores, cannot be fully exercised in the current single-node sandbox. Although \sys{} may generate structurally faithful code, performance claims involving throughput, tail latency, partition behavior, or hardware offloading remain unverifiable within the framework.
3) \textit{Extensions to existing codebases.} Papers whose contributions are implemented as modifications to large external projects, such as Kubernetes scheduler hooks or LLVM passes, are not fully supported. \sys{} currently does not clone, pin, and patch the required upstream codebase. As a result, it may generate self-contained code that is structurally plausible but cannot compile against the intended project. This gap motivates future work on leveraging existing crawled codebases.}

\subsection{Future Work}

\paragraph{Dynamic architecture refinement.}
\emph{ARA} currently freezes the architecture once \texttt{arch.json} is generated. In future work, runtime evidence such as repeated \texttt{ModuleNotFoundError}s, persistent interface type mismatches, or consistently empty module outputs can be used to revise the architecture. We will add a lightweight re-entry point that re-prompts the architecture-design agent with accumulated failure logs, allowing it to propose module merges, splits, or interface revisions without restarting the full pipeline.

\paragraph{Leveraging existing crawled code.}
Many systems papers are more naturally reproduced by modifying an existing open-source repository than by generating code from scratch. We will augment the paper-parsing stage with optional repository URLs, clone and pin the target revision, and guide code generation to emit \texttt{<patch>} blocks against the cloned tree. Existing mechanisms for unique patch matching and safe path resolution can support this workflow, but they must be integrated with code search to identify the appropriate upstream files for modification.

\paragraph{Flexible sandbox construction.}
Current sandbox construction has limited support for code that depends on specific kernels, compilers, or hardware toolchains. To address this, we will augment the architecture agent with a custom \texttt{docker-compose.yml} or a Dockerfile with a non-default base image. The sandbox can then be provisioned from this explicit specification rather than from a heuristic runtime profile, enabling more environment-coupled artifacts to be built and tested while requiring additional validation of model-generated container specifications.

\paragraph{Shared failure memory.}
The current audit and runtime-repair steps are stateless across steps: once a repair attempt terminates, its failure trace is logged but not reused. We will introduce a per-paper failure memory keyed by paper, module, step, and error category. This memory can be consulted at the beginning of each repair pass to avoid repeated failed edits, such as \texttt{NOT\_FOUND} or ambiguous \texttt{<old>} matches, and can provide prior successful fixes as in-context examples. This lightweight mechanism would allow the agent to learn from prior failures within the same reproduction task.
\section{Conclusion}
\label{sec:conclusion}
In this paper, we present \sys, a multi-agent framework for reproducing network research results from academic papers. \sys addresses the main barriers in this task: long and unstructured paper content, complex system logic, cross-file dependencies, and the lack of reliable verification. Instead of asking a single LLM to generate an entire codebase at once, \sys divides the workflow into content parsing, architecture design, code generation, and audit\&repair. Its \textit{Shared Memory} keeps paper information, interface definitions, intermediate states, and generated code in a consistent form across agents. Together with schema-first generation, structured reasoning, static analysis, and sandbox-based runtime feedback, \sys improves both executability and semantic alignment. Experiments on representative networking papers show that \sys produces more reliable reproductions than existing baselines while using fewer unnecessary tokens. More broadly, we hope \sys can motivate future work that further explores how LLM-based systems can support network research, not only in reproduction, but also in system design, validation, evaluation, and artifact generation.

\vspace{2em}
\section*{Ethical Statement}
\xyx{This work does not raise any ethical issues.}
\vspace{1em}

\begin{acks}
We thank the anonymous reviewers for their valuable comments. This work is funded by the National Key R\&D Program of China (Grant No. 2024YFB2906900), the National Natural Science Foundation of China (Grant Nos. 62532010, 62572408, and 62522203), and the Fujian Provincial Science and Technology Project (Grant No. 2025H6021).
\end{acks}

\clearpage
\bibliographystyle{ACM-Reference-Format}
\bibliography{reference}

@inproceedings{jin2018netchain,
author = {Jin, Xin and Li, Xiaozhou and Zhang, Haoyu and Foster, Nate and Lee, Jeongkeun and Soul{\'e}, Robert and Kim, Changhoon and Stoica, Ion},
title = {$\{$NetChain$\}$:$\{$Scale-Free$\}$$\{$Sub-RTT$\}$ coordination},
booktitle = {Proceedings of the 15th USENIX Symposium on Networked Systems Design and Implementation},
series = {NSDI '18},
year = {2018},
month = apr,
location = {Renton, WA, USA},
pages = {35--49},
numpages = {15},
isbn = {978-1-939133-01-4},
url = {https://www.usenix.org/conference/nsdi18/presentation/jin},
publisher = {USENIX Association},
address = {Berkeley, CA, USA},
}

@inproceedings{xu2023teal,
author = {Xu, Zhiying and Yan, Francis Y and Singh, Rachee and Chiu, Justin T and Rush, Alexander M and Yu, Minlan},
title = {Teal: Learning-Accelerated Optimization of WAN Traffic Engineering},
booktitle = {Proceedings of the ACM SIGCOMM 2023 Conference},
series = {SIGCOMM '23},
year = {2023},
month = sep,
location = {New York, NY, USA},
pages = {378--393},
numpages = {16},
doi = {10.1145/3603269.3604857},
url = {https://dl.acm.org/doi/10.1145/3603269.3604857},
acmid = {3604857},
isbn = {979-8-4007-0236-5},
publisher = {ACM},
address = {New York, NY, USA},
}

@inproceedings{steffen2020probabilistic,
author = {Steffen, Samuel and Gehr, Timon and Tsankov, Petar and Vanbever, Laurent and Vechev, Martin},
title = {Probabilistic Verification of Network Configurations},
booktitle = {Proceedings of the 2020 Annual Conference of the ACM Special Interest Group on Data Communication on the Applications, Technologies, Architectures, and Protocols for Computer Communication},
series = {SIGCOMM '20},
year = {2020},
month = aug,
location = {New York City, NY, USA (Virtual)},
pages = {750--764},
numpages = {15},
doi = {10.1145/3387514.3405900},
url = {https://dl.acm.org/doi/10.1145/3387514.3405900},
acmid = {3405900},
isbn = {978-1-4503-7955-7},
publisher = {ACM},
address = {New York, NY, USA},
}

@inproceedings{kakarla2020finding,
author = {Kakarla, Siva Kesava Reddy and Tang, Alan and Beckett, Ryan and Jayaraman, Karthick and Millstein, Todd and Tamir, Yuval and Varghese, George},
title = {Finding Network Misconfigurations by Automatic Template Inference},
booktitle = {Proceedings of the 17th USENIX Symposium on Networked Systems Design and Implementation},
series = {NSDI '20},
year = {2020},
month = feb,
location = {Santa Clara, CA, USA},
pages = {999--1013},
numpages = {15},
isbn = {978-1-939133-17-5},
url = {https://www.usenix.org/conference/nsdi20/presentation/kakarla},
publisher = {USENIX Association},
address = {Berkeley, CA, USA},
}

@inproceedings{shi2023large,
    title = 	 {Large Language Models Can Be Easily Distracted by Irrelevant Context},
    author =       {Shi, Freda and Chen, Xinyun and Misra, Kanishka and Scales, Nathan and Dohan, David and Chi, Ed H. and Sch\"{a}rli, Nathanael and Zhou, Denny},
    booktitle = 	 {Proceedings of the 40th International Conference on Machine Learning},
    pages = 	 {31210--31227},
    year = 	 {2023},
    editor = 	 {Krause, Andreas and Brunskill, Emma and Cho, Kyunghyun and Engelhardt, Barbara and Sabato, Sivan and Scarlett, Jonathan},
    volume = 	 {202},
    series = 	 {Proceedings of Machine Learning Research},
    month = 	 {23--29 Jul},
    publisher =    {PMLR},
    url = 	 {https://proceedings.mlr.press/v202/shi23a.html}
}

@inproceedings{xiang2023toward,
author = {Xiang, Qiao and Lin, Yuling and Fang, Mingjun and Huang, Bang and Huang, Siyong and Wen, Ridi and Le, Franck and Kong, Linghe and Shu, Jiwu},
title = {Toward Reproducing Network Research Results Using Large Language Models},
year = {2023},
isbn = {9798400704154},
publisher = {Association for Computing Machinery},
address = {New York, NY, USA},
url = {https://doi.org/10.1145/3626111.3628189},
doi = {10.1145/3626111.3628189},
booktitle = {Proceedings of the 22nd ACM Workshop on Hot Topics in Networks},
pages = {56–62},
numpages = {7},
location = {Cambridge, MA, USA},
series = {HotNets '23}
}

@misc{github_copilot,
  title        = {GitHub Copilot},
  author       = {{GitHub}},
  year         = {2022},
  howpublished = {\url{https://github.com/features/copilot/}},
  note         = {Accessed: 2025-09}
}

@misc{openai_chatgpt,
  title        = {ChatGPT},
  author       = {{OpenAI}},
  year         = {2022},
  howpublished = {\url{https://chat.openai.com/}},
  note         = {Accessed: 2025-09}
}

@article{paszke2019pytorch,
author = {Paszke, Adam and
    Gross, Sam and
    Massa, Francisco and
    Lerer, Adam and
    Bradbury, James and
    Chanan, Gregory and
    Killeen, Trevor and
    Lin, Zeming and
    Gimelshein, Natalia and
    Antiga, Luca and
    Desmaison, Alban and
    K{\"o}pf, Andreas and
    Yang, Edward and
    DeVito, Zachary and
    Raison, Martin and
    Tejani, Alykhan and
    Chilamkurthy, Sasank and
    Steiner, Benoit and
    Fang, Lu and
    Bai, Junjie and
    Chintala, Soumith},
title = {Pytorch: An imperative style, high-performance deep learning library},
journal = {Advances in Neural Information Processing Systems},
issue_date = {2019},
volume = {32},
year = {2019},
pages = {8026--8037},
numpages = {12},
url = {https://proceedings.neurips.cc/paper_files/paper/2019/hash/bdbca288fee7f92f2bfa9f7012727740-Abstract.html},
doi = {10.5555/3454287.3455008},
publisher = {Curran Associates, Inc.},
address = {Red Hook, NY, USA},
}

@misc{cursor,
  title        = {Cursor: An AI-Powered Code Editor},
  author       = {{Anysphere}},
  year         = {2024},
  howpublished = {\url{https://cursor.com/}},
  note         = {Accessed: 2025-09}
}

@article{zhang2024pydex,
    author = {Zhang, Jialu and Cambronero, Jos\'{e} Pablo and Gulwani, Sumit and Le, Vu and Piskac, Ruzica and Soares, Gustavo and Verbruggen, Gust},
    title = {PyDex: Repairing Bugs in Introductory Python Assignments using LLMs},
    year = {2024},
    issue_date = {April 2024},
    publisher = {Association for Computing Machinery},
    address = {New York, NY, USA},
    volume = {8},
    number = {OOPSLA1},
    url = {https://doi.org/10.1145/3649850},
    doi = {10.1145/3649850},
    journal = {Proc. ACM Program. Lang.},
    month = apr,
    articleno = {133},
    numpages = {25},
}

@article{rahmani2021multi,
author = {Rahmani, Kia and Raza, Mohammad and Gulwani, Sumit and Le, Vu and Morris, Daniel and Radhakrishna, Arjun and Soares, Gustavo and Tiwari, Ashish},
title = {Multi-modal program inference: a marriage of pre-trained language models and component-based synthesis},
journal = {Proceedings of the ACM on Programming Languages},
issue_date = {October 2021},
volume = {5},
number = {OOPSLA},
month = oct,
year = {2021},
issn = {2475-1421},
pages = {1--29},
numpages = {29},
url = {https://dl.acm.org/doi/10.1145/3485535},
doi = {10.1145/3485535},
acmid = {3485535},
publisher = {ACM},
address = {New York, NY, USA},
}

@inproceedings{yen2021semi,
author = {Yen, Jane and L{\'e}vai, Tam{\'a}s and Ye, Qinyuan and Ren, Xiang and Govindan, Ramesh and Raghavan, Barath},
title = {Semi-Automated Protocol Disambiguation and Code Generation},
booktitle = {Proceedings of the 2021 ACM SIGCOMM Conference},
series = {SIGCOMM '21},
year = {2021},
month = aug,
location = {Virtual Event, USA},
pages = {272--286},
numpages = {15},
doi = {10.1145/3452296.3472910},
url = {https://dl.acm.org/doi/10.1145/3452296.3472910},
acmid = {3472910},
isbn = {978-1-4503-8629-6},
publisher = {ACM},
address = {New York, NY, USA},
}

@article{seo2025paper2code,
  title={Paper2code: Automating code generation from scientific papers in machine learning},
  author={Seo, Minju and Baek, Jinheon and Lee, Seongyun and Hwang, Sung Ju},
  journal={arXiv preprint arXiv:2504.17192},
  year={2025}
}

@article{luo2025intention,
author = {Luo, Yi and Shi, Linghang and Li, Yihao and Zhuang, Aobo and Gong, Yeyun and Liu, Ling and Lin, Chen},
title = {From intention to implementation: automating biomedical research via LLMs},
journal = {Science China Information Sciences},
issue_date = {2025},
volume = {68},
number = {7},
year = {2025},
pages = {78--95},
numpages = {18},
url ={http://scis.scichina.com/en/2025/170105.pdf},
doi = {10.1007/s11432-024-4485-0},
}

@article{lin2025autop2c,
  title={AutoP2C: An LLM-Based Agent Framework for Code Repository Generation from Multimodal Content in Academic Papers},
  author={Lin, Zijie and Shen, Yiqing and Cai, Qilin and Sun, He and Zhou, Jinrui and Xiao, Mingjun},
  journal={arXiv preprint arXiv:2504.20115},
  year={2025}
}

@inproceedings{wang2025llm,
    author = {Wang, Yibo and Hou, Yunan and Lai, Zeqi and Li, Hewu and Wu, Qian and Liu, Jun and Li, Yuanjie and Xie, Xin and Han, Zhifeng},
    title = {How LLM Saved Me from Struggling with Experiment Reproduction: LEO Networking as A Case Study},
    year = {2025},
    isbn = {9798400720901},
    publisher = {Association for Computing Machinery},
    address = {New York, NY, USA},
    url = {https://doi.org/10.1145/3748749.3749084},
    doi = {10.1145/3748749.3749084},
    booktitle = {Proceedings of the 2025 3rd Workshop on LEO Networking and Communication},
    pages = {1–-7},
    numpages = {7},
    location = {Coimbra, Portugal},
    series = {LEO-NET '25}
}

@inproceedings{chen2022software,
    author = {Chen, Huangxun and Miao, Yukai and Chen, Li and Sun, Haifeng and Xu, Hong and Liu, Libin and Zhang, Gong and Wang, Wei},
    title = {Software-defined network assimilation: bridging the last mile towards centralized network configuration management with NAssim},
    year = {2022},
    isbn = {9781450394208},
    publisher = {Association for Computing Machinery},
    address = {New York, NY, USA},
    url = {https://doi.org/10.1145/3544216.3544244},
    doi = {10.1145/3544216.3544244},
    booktitle = {Proceedings of the ACM SIGCOMM 2022 Conference},
    pages = {281–297},
    numpages = {17},
    location = {Amsterdam, Netherlands},
    series = {SIGCOMM '22}
}

@inproceedings{abuzaid2021contracting,
    author = {Firas Abuzaid and Srikanth Kandula and Behnaz Arzani and Ishai Menache and Matei Zaharia and Peter Bailis},
    title = {Contracting Wide-area Network Topologies to Solve Flow Problems Quickly},
    booktitle = {Proceedings of the 18th USENIX Symposium on Networked Systems Design and Implementation},
    series = {NSDI '21},
    year = {2021},
    month = apr,
    location = {Boston, MA, USA (Virtual)},
    pages = {175--200},
    numpages = {26},
    isbn = {978-1-939133-21-2},
    url = {https://www.usenix.org/conference/nsdi21/presentation/abuzaid},
    publisher = {USENIX Association},
    address = {Berkeley, CA, USA},
}

@inproceedings{kakarla2020otot,
    author = {Kakarla, Siva Kesava Reddy and Beckett, Ryan and Arzani, Behnaz and Millstein, Todd and Varghese, George},
    title = {GRooT: Proactive Verification of DNS Configurations},
    year = {2020},
    isbn = {9781450379557},
    publisher = {Association for Computing Machinery},
    address = {New York, NY, USA},
    url = {https://doi.org/10.1145/3387514.3405871},
    doi = {10.1145/3387514.3405871},
    booktitle = {Proceedings of the Annual Conference of the ACM Special Interest Group on Data Communication on the Applications, Technologies, Architectures, and Protocols for Computer Communication},
    pages = {310–328},
    numpages = {19},
    location = {Virtual Event, USA},
    series = {SIGCOMM '20}
}

@article{yan2017learning,
author = {Yan, Lisa and McKeown, Nick},
title = {Learning Networking by Reproducing Research Results},
year = {2017},
issue_date = {April 2017},
publisher = {Association for Computing Machinery},
address = {New York, NY, USA},
volume = {47},
number = {2},
issn = {0146-4833},
url = {https://doi.org/10.1145/3089262.3089266},
doi = {10.1145/3089262.3089266},
journal = {SIGCOMM Comput. Commun. Rev.},
month = may,
pages = {19–26},
numpages = {8},
}

@article{yang2015real,
author = {Yang, Hongkun and Lam, Simon S.},
title = {Real-Time Verification of Network Properties Using Atomic Predicates},
journal = {IEEE/ACM Transactions on Networking},
issue_date = {April 2016},
volume = {24},
number = {2},
month = apr,
year = {2016},
issn = {1063-6692},
pages = {887--900},
numpages = {14},
url = {https://doi.org/10.1109/TNET.2015.2398197},
doi = {10.1109/TNET.2015.2398197},
publisher = {IEEE},
address = {New York, NY, USA},
}

@article{li2025structured,
  title = {Structured Chain-of-Thought Prompting for Code Generation},
  author = {Li, Jia and Li, Ge and Li, Yongmin and Jin, Zhi},
  journal = {arXiv preprint arXiv:2305.06599},
  year = {2023},
  url ={https://arxiv.org/abs/2305.06599}
}

@inproceedings{zhang2020apkeep,
author = {Zhang, Peng and Liu, Xu and Yang, Hongkun and Kang, Ning and Gu, Zhengchang and Li, Hao},
title = {$\{$APKeep$\}$: Realtime Verification for Real Networks},
booktitle = {Proceedings of the 17th USENIX Symposium on Networked Systems Design and Implementation},
series = {NSDI '20},
year = {2020},
month = feb,
location = {Santa Clara, CA, USA},
pages = {241--255},
numpages = {15},
isbn = {978-1-939133-13-7},
url = {https://www.usenix.org/conference/nsdi20/presentation/zhang-peng},
publisher = {USENIX Association},
address = {Berkeley, CA, USA},
}

@inproceedings{handigol2012reproducible,
author = {Handigol, Nikhil and Heller, Brandon and Jeyakumar, Vimalkumar and Lantz, Bob and McKeown, Nick},
title = {Reproducible network experiments using container-based emulation},
year = {2012},
isbn = {9781450317757},
publisher = {Association for Computing Machinery},
address = {New York, NY, USA},
url = {https://doi.org/10.1145/2413176.2413206},
doi = {10.1145/2413176.2413206},
booktitle = {Proceedings of the 8th International Conference on Emerging Networking Experiments and Technologies},
pages = {253–264},
numpages = {12},
location = {Nice, France},
series = {CoNEXT '12}
}

@article{li2025deepcode,
  title={DeepCode: Open Agentic Coding},
  author={Li, Zongwei and Li, Zhonghang and Guo, Zirui and Ren, Xubin and Huang, Chao},
  journal={arXiv preprint arXiv:2512.07921},
  year={2025}
}

@inproceedings{zhao2025autoreproduce,
  author = {Zhao, Xuanle and Sang, Zilin and Li, Yuxuan and Shi, Qi and Zhao, Weilun and Wang, Shuo and Zhang, Duzhen and Han, Xu and Liu, Zhiyuan and Sun, Maosong},
  title = {{AutoReproduce}: Automatic {AI} Experiment Reproduction with Paper Lineage},
  booktitle = {Proceedings of the 64th Annual Meeting of the Association for Computational Linguistics (Volume 1: Long Papers)},
  year = {2026},
  month = jul,
  address = {San Diego, California, USA},
  pages = {21920--21942},
  doi = {10.18653/v1/2026.acl-long.1001},
  url = {https://aclanthology.org/2026.acl-long.1001/},
  publisher = {Association for Computational Linguistics}
}

@article{wu2024autogen,
  title={Autogen: Enabling next-gen llm applications via multi-agent conversation},
  author={Wu, Qingyun and
    Bansal, Gagan and
    Zhang, Jieyu and
    Wu, Yiran and
    Li, Beibin and
    Zhu, Erkang and
    Jiang, Li and
    Zhang, Xiaoyun and
    Zhang, Shaokun and
    Liu, Jiale and
    Awadallah, Ahmed Hassan and
    White, Ryen W. and
    Burger, Doug and
    Wang, Chi},
  journal={arXiv preprint arXiv:2308.08155},
  year={2023}
}

@online{claude_context_windows,
  author = {{Anthropic}},
  title  = {{Build with Claude: Context Windows}},
  url    = {https://platform.claude.com/docs/en/build-with-claude/context-windows},
  note   = {Accessed: 2026-07-03}
}

@misc{claude,
	howpublished = {\url{https://claude.com/product/claude-code}},
	title = {{Claude Code - AI coding agent for terminal \& IDE | Claude}},
    author = {{Anthropic}},
    year = {2025}
}

@article{gao2023retrieval,
  title={Retrieval-augmented generation for large language models: A survey},
  author={Gao, Yunfan and Xiong, Yun and Gao, Xinyu and Jia, Kangxiang and Pan, Jinliu and Bi, Yuxi and Dai, Yi and Sun, Jiawei and Wang, Meng and Wang, Haofen},
  journal={arXiv preprint arXiv:2312.10997},
  year={2023}
}

@article{liu2024lost,
author = {Liu, Nelson F. and Lin, Kevin and Hewitt, John and Paranjape, Ashwin and Bevilacqua, Michele and Petroni, Fabio and Liang, Percy},
title = {Lost in the Middle: How Language Models Use Long Contexts},
journal = {Transactions of the Association for Computational Linguistics},
issue_date = {2024},
volume = {12},
year = {2024},
pages = {157--173},
numpages = {17},
url = {https://aclanthology.org/2024.tacl-1.9/},
doi = {10.1162/tacl_a_00638},
publisher = {MIT Press},
address = {Cambridge, MA},
}

@inproceedings{blecher2023nougat,
author = {Blecher, Lukas and Cucurull Preixens, Guillem and Scialom, Thomas and Stojnic, Robert},
title = {Nougat: Neural Optical Understanding for Academic Documents},
booktitle = {Proceedings of the 12th International Conference on Learning Representations},
series = {ICLR '24},
year = {2024},
month = may,
location = {Vienna, Austria},
pages = {37646--37663},
numpages = {18},
volume = {2024},
url = {https://proceedings.iclr.cc/paper_files/paper/2024/file/a39a9aceda771cded859ae7560530e09-Paper-Conference.pdf},
publisher = {ICLR},
}

@article{wang2024mineru,
  title={Mineru: An open-source solution for precise document content extraction},
  author={Wang, Bin and
    Xu, Chao and
    Zhao, Xiaomeng and
    Ouyang, Linke and
    Wu, Fan and
    Zhao, Zhiyuan and
    Xu, Rui and
    Liu, Kaiwen and
    Qu, Yuan and
    Shang, Fukai and
    Zhang, Bo and
    Wei, Liqun and
    Sui, Zhihao and
    Li, Wei and
    Shi, Botian and
    Qiao, Yu and
    Lin, Dahua and
    He, Conghui},
  journal={arXiv preprint arXiv:2409.18839},
  year={2024}
}

@article{tian2024scicode,
author = {Tian, Minyang and
    Gao, Luyu and
    Zhang, Shizhuo Dylan and
    Chen, Xinan and
    Fan, Cunwei and
    Guo, Xuefei and
    Haas, Roland and
    Ji, Pan and
    Krongchon, Kittithat and
    Li, Yao and
    Liu, Shengyan and
    Luo, Di and
    Ma, Yutao and
    Tong, Hao and
    Trinh, Kha and
    Tian, Chenyu and
    Wang, Zihan and
    Wu, Bohao and
    Xiong, Yanyu and
    Yin, Shengzhu and
    Zhu, Minhui and
    Lieret, Kilian and
    Lu, Yanxin and
    Liu, Genglin and
    Du, Yufeng and
    Tao, Tianhua and
    Press, Ofir and
    Callan, Jamie and
    Huerta, Eliu and
    Peng, Hao},
title = {SciCode: A Research Coding Benchmark Curated by Scientists},
journal = {Advances in Neural Information Processing Systems},
issue_date = {December 2024},
volume = {37},
year = {2024},
pages = {30624--30650},
numpages = {27},
url = {https://proceedings.neurips.cc/paper_files/paper/2024/hash/330122077c40f3408488152702356766-Abstract.html},
doi = {10.5555/3690531.3699047},
acmid = {3699047},
publisher = {Curran Associates, Inc.},
address = {Red Hook, NY, USA},
}

@inproceedings{jimenez2023swe,
author = {Jimenez, Carlos E and Yang, John and Wettig, Alexander and Yao, Shunyu and Pei, Kexin and Press, Ofir and Narasimhan, Karthik},
title = {SWE-bench: Can Language Models Resolve Real-world Github Issues?},
booktitle = {Proceedings of the 12th International Conference on Learning Representations},
series = {ICLR '24},
year = {2024},
month = may,
location = {Vienna, Austria},
pages = {54107--54157},
numpages = {51},
volume = {2024},
url = {https://proceedings.iclr.cc/paper_files/paper/2024/file/edac78c3e300629acfe6cbe9ca88fb84-Paper-Conference.pdf},
publisher = {ICLR},
}

@article{zelikman2023parsel,
author = {Zelikman, Eric and Huang, Qian and Poesia, Gabriel and Goodman, Noah and Haber, Nick},
title = {Parsel: Algorithmic Reasoning with Language Models by Composing Decompositions},
journal = {Advances in Neural Information Processing Systems},
issue_date = {December 2023},
volume = {36},
year = {2023},
pages = {31466--31523},
numpages = {58},
url = {https://proceedings.neurips.cc/paper_files/paper/2023/hash/2711f35a451174733f7081802f283773-Abstract.html},
doi = {10.5555/3666122.3668746},
acmid = {3668746},
publisher = {Curran Associates, Inc.},
address = {Red Hook, NY, USA},
}

@article{bairi2024codeplan,
  author = {Bairi, Ramakrishna and Sonwane, Atharv and Kanade, Aditya and C., Vageesh D. and Iyer, Arun and Parthasarathy, Suresh and Rajamani, Sriram and Ashok, B. and Shet, Shashank},
  title = {CodePlan: Repository-Level Coding using LLMs and Planning},
  journal = {Proceedings of the ACM on Software Engineering},
  volume = {1},
  number = {FSE},
  year = {2024},
  month = jul,
  issue_date = {July 2024},
  pages = {675--698},
  articleno = {31},
  numpages = {24},
  publisher = {Association for Computing Machinery},
  address = {New York, NY, USA},
  doi = {10.1145/3643757},
  url = {https://doi.org/10.1145/3643757},
}

@inproceedings{qian2024chatdev,
author = {Qian, Chen and Liu, Wei and Liu, Hongzhang and Chen, Nuo and Dang, Yufan and Li, Jiahao and Yang, Cheng and Chen, Weize and Su, Yusheng and Cong, Xin and Xu, Juyuan and Li, Dahai and Liu, Zhiyuan and Sun, Maosong},
title = {ChatDev: Communicative Agents for Software Development},
booktitle = {Proceedings of the 62nd Annual Meeting of the Association for Computational Linguistics (Volume 1: Long Papers)},
series = {ACL '24},
year = {2024},
month = aug,
location = {Bangkok, Thailand},
pages = {15174--15186},
numpages = {13},
doi = {10.18653/v1/2024.acl-long.810},
url = {https://aclanthology.org/2024.acl-long.810/},
publisher = {Association for Computational Linguistics},
address = {Stroudsburg, PA, USA},
}

@inproceedings{hong2023metagpt,
author = {Hong, Sirui and Zhuge, Mingchen and Chen, Jonathan and Zheng, Xiawu and Cheng, Yuheng and Wang, Jinlin and Zhang, Ceyao and Wang, Zili and Yau, Steven and Lin, Zijuan and Zhou, Liyang and Ran, Chenyu and Xiao, Lingfeng and Wu, Chenglin and Schmidhuber, J\"{u}rgen},
title = {MetaGPT: Meta Programming for A Multi-Agent Collaborative Framework},
booktitle = {Proceedings of the 12th International Conference on Learning Representations},
series = {ICLR '24},
year = {2024},
month = may,
location = {Vienna, Austria},
pages = {23247--23275},
numpages = {29},
volume = {2024},
url = {https://proceedings.iclr.cc/paper_files/paper/2024/file/6507b115562bb0a305f1958ccc87355a-Paper-Conference.pdf},
publisher = {ICLR},
}

@inproceedings{zhang2023repocoder,
author = {Zhang, Fengji and Chen, Bei and Zhang, Yue and Keung, Jacky and Liu, Jin and Zan, Daoguang and Mao, Yi and Lou, Jian-Guang and Chen, Weizhu},
title = {RepoCoder: Repository-Level Code Completion Through Iterative Retrieval and Generation},
booktitle = {Proceedings of the 2023 Conference on Empirical Methods in Natural Language Processing},
series = {EMNLP '23},
year = {2023},
month = dec,
location = {Singapore},
pages = {2471--2484},
numpages = {14},
doi = {10.18653/v1/2023.emnlp-main.152},
url = {https://aclanthology.org/2023.emnlp-main.152/},
publisher = {Association for Computational Linguistics},
address = {Stroudsburg, PA, USA},
}

@article{poesia2022synchromesh,
  title={Synchromesh: Reliable code generation from pre-trained language models},
  author={Poesia, Gabriel and Polozov, Oleksandr and Le, Vu and Tiwari, Ashish and Soares, Gustavo and Meek, Christopher and Gulwani, Sumit},
  journal={arXiv preprint arXiv:2201.11227},
  year={2022}
}

@article{shinn2023reflexion,
author = {Shinn, Noah and Cassano, Federico and Gopinath, Ashwin and Narasimhan, Karthik and Yao, Shunyu},
title = {Reflexion: Language Agents with Verbal Reinforcement Learning},
journal = {Advances in Neural Information Processing Systems},
issue_date = {December 2023},
volume = {36},
year = {2023},
pages = {8634--8652},
numpages = {19},
url = {https://dl.acm.org/doi/10.5555/3666122.3666499},
doi = {10.5555/3666122.3666499},
acmid = {3666499},
publisher = {Curran Associates, Inc.},
address = {Red Hook, NY, USA},
}

@inproceedings{yang2024practical,
author = {Yang, Lei and Gilad, Yossi and Alizadeh, Mohammad},
title = {Practical Rateless Set Reconciliation},
booktitle = {Proceedings of the ACM SIGCOMM 2024 Conference},
series = {SIGCOMM '24},
year = {2024},
month = aug,
location = {Sydney, NSW, Australia},
pages = {595--612},
numpages = {18},
doi = {10.1145/3651890.3672219},
url = {https://dl.acm.org/doi/10.1145/3651890.3672219},
acmid = {3672219},
isbn = {979-8-4007-0614-1},
publisher = {ACM},
address = {New York, NY, USA},
}

@misc{ruff,
  author = {{Astral}},
  title = {{Ruff: An extremely fast Python linter and code formatter}},
  howpublished = {\url{https://github.com/astral-sh/ruff}},
  year = {2026}
}

@inproceedings{liu2025rtadev,
  author = {Liu, Jie and Wang, Guohua and Yang, Ronghui and Zeng, Jiajie and Zhao, Mengchen and Cai, Yi},
  title = {{RTADev}: Intention Aligned Multi-Agent Framework for Software Development},
  booktitle = {Findings of the Association for Computational Linguistics: ACL 2025},
  year = {2025},
  month = jul,
  address  = {Vienna, Austria},
  pages  = {1548--1581},
  publisher = {Association for Computational Linguistics},
  doi = {10.18653/v1/2025.findings-acl.80},
  url= {https://aclanthology.org/2025.findings-acl.80/}
}

@inproceedings{ding2023crosscodeeval,
author = {Ding, Yangruibo and Wang, Zijian and Ahmad, Wasi Uddin and Ding, Hantian and Tan, Ming and Jain, Nihal and Ramanathan, Murali Krishna and Nallapati, Ramesh and Bhatia, Parminder and Roth, Dan and Xiang, Bing},
title = {CROSSCODEEVAL: A Diverse and Multilingual Benchmark for Cross-File Code Completion},
booktitle = {Proceedings of the 37th International Conference on Neural Information Processing Systems},
series = {NIPS '23},
year = {2023},
month = dec,
location = {New Orleans, LA, USA},
pages = {46701--46723},
numpages = {23},
volume = {36},
url = {https://proceedings.neurips.cc/paper_files/paper/2023/hash/8162f6586e88c3e8575f39e345473092-Paper-Conference.pdf},
doi = {10.5555/3666122.3668145},
acmid = {3668145},
publisher = {Curran Associates Inc.},
address = {Red Hook, NY, USA},
editor = {Alice Oh and Tristan Naumann and Amir Globerson and Kate Saenko and Moritz Hardt and Sergey Levine},
isbn = {9781713899921},
}

@inproceedings{chen2023teaching,
author = {Chen, Xinyun and Lin, Maxwell and Schaerli, Nathanael and Zhou, Denny},
title = {Teaching Large Language Models to Self-Debug},
booktitle = {Proceedings of the 12th International Conference on Learning Representations},
series = {ICLR '24},
year = {2024},
month = may,
location = {Vienna, Austria},
pages = {8746--8825},
numpages = {80},
volume = {2024},
url = {https://proceedings.iclr.cc/paper_files/paper/2024/file/2460396f2d0d421885997dd1612ac56b-Paper-Conference.pdf},
publisher = {ICLR},
}

@misc{gulati2025controllable,
  author={Gulati, Aryan},
  title={Controllable LLM Debugging: Knowing when to Stop Matters},
  year={2024},
  url ={https://cs191.stanford.edu/projects/Gulati,%20Aryan_NLP%20191W.pdf},
  note = {Stanford CS191/NLP 191W course project report. Accessed: 2026-07-03}
}

@article{yang2023intercode,
author = {Yang, John and Prabhakar, Akshara and Narasimhan, Karthik and Yao, Shunyu},
title = {Intercode: Standardizing and Benchmarking Interactive Coding with Execution Feedback},
journal = {Advances in Neural Information Processing Systems},
issue_date = {December 2023},
volume = {36},
year = {2023},
pages = {23826--23854},
numpages = {29},
url = {https://proceedings.neurips.cc/paper_files/paper/2023/hash/906c74d4f45430c1f33f36350598446c-Abstract.html},
doi = {10.5555/3666122.3668014},
acmid = {3668014},
publisher = {Curran Associates, Inc.},
address = {Red Hook, NY, USA},
}

@inproceedings{wang2025characteristics,
author = {Wang, Zhijie and Zhou, Zijie and Song, Da and Huang, Yuheng and Chen, Shengmai and Ma, Lei and Zhang, Tianyi},
title = {Towards Understanding the Characteristics of Code Generation Errors Made by Large Language Models},
booktitle = {Proceedings of the 47th IEEE/ACM International Conference on Software Engineering},
series = {ICSE '25},
year = {2025},
location = {Ottawa, ON, Canada},
pages = {2587--2599},
numpages = {13},
isbn = {979-8-3315-0569-1},
doi = {10.1109/ICSE55347.2025.00180},
url = {https://dl.acm.org/doi/10.1109/ICSE55347.2025.00180},
publisher = {IEEE},
address = {Piscataway, NJ, USA},
}

@inproceedings{ni2023lever,
author = {Ni, Ansong and Iyer, Srini and Radev, Dragomir and Stoyanov, Ves and Yih, Wen-tau and Wang, Sida I and Lin, Xi Victoria},
title = {LEVER: Learning to Verify Language-to-Code Generation with Execution},
booktitle = {Proceedings of the 40th International Conference on Machine Learning},
series = {ICML '23},
year = {2023},
month = jul,
location = {Honolulu, Hawaii, USA},
pages = {26106--26128},
numpages = {23},
volume = {202},
url = {https://proceedings.mlr.press/v202/ni23/ni23.pdf},
publisher = {PMLR},
address = {Cambridge, MA, USA},
}

@misc{langchain,
  author       = {{LangChain AI}},
  title        = {LangChain: Building applications with LLMs through composability},
  year         = {2022},
  howpublished = {\url{https://github.com/langchain-ai/langchain}},
  note         = {Accessed: 2026-02-06}
}

@misc{docker,
  title        = {{Docker}: Lightweight Linux Containers for Consistent Development and Operations},
  author       = {{Docker Inc.}},
  year         = {2013},
  url          = {https://www.docker.com/},
  note         = {Accessed: 2026-02-07}
}

@inproceedings{liu2023repobench,
author = {Liu, Tianyang and Xu, Canwen and McAuley, Julian},
title = {RepoBench: Benchmarking Repository-Level Code Auto-Completion Systems},
booktitle = {Proceedings of the 12th International Conference on Learning Representations},
series = {ICLR '24},
year = {2024},
month = may,
location = {Vienna, Austria},
pages = {47832--47850},
numpages = {19},
volume = {2024},
url = {https://proceedings.iclr.cc/paper_files/paper/2024/file/893699c6824d4423b4092d4f898d00c7-Paper-Conference.pdf},
publisher = {ICLR},
}

@inproceedings{abadi2016tensorflow,
author = {Abadi, Mart\'{\i}n and Barham, Paul and Chen, Jianmin and Chen, Zhifeng and Davis, Andy and Dean, Jeffrey and Devin, Matthieu and Ghemawat, Sanjay and Irving, Geoffrey and Isard, Michael and Kudlur, Manjunath and Levenberg, Josh and Monga, Rajat and Moore, Sherry and Murray, Derek G. and Steiner, Benoit and Tucker, Paul and Vasudevan, Vijay and Warden, Pete and Wicke, Martin and Yu, Yuan and Zheng, Xiaoqiang},
title = {TensorFlow: a system for large-scale machine learning},
year = {2016},
isbn = {9781931971331},
publisher = {USENIX Association},
address = {USA},
booktitle = {Proceedings of the 12th USENIX Conference on Operating Systems Design and Implementation},
pages = {265–283},
numpages = {19},
location = {Savannah, GA, USA},
series = {OSDI'16}
}
\appendix
\newpage
\textbf{Note:} Appendices are supporting material that has not been peer-reviewed.
\section{Appendix}

\subsection{Detailed Analysis of Reproduction Performance Results} \label{sec:appd_reproduction_analysis}

\subsubsection{NetChain}
We evaluate the end-to-end reproducibility of NetChain by testing whether different baselines, given identical inputs (paper, topology files, and command traces), can generate all four required artifacts (P4 data plane, Python control plane, client, and orchestration scripts) and successfully execute on a BMv2-based system.
\begin{itemize}
\item \textbf{RepLLM.}
Across the automatically generated RepLLM versions, the reconstruction progressively covers the four-module NetChain architecture, including the P4 data plane, control-plane logic, client-side traffic generation, and orchestration scripts. The generated artifacts differ in the degree of client completeness, failure-handling logic, topology integration, and execution-environment alignment, but they collectively recover the main structural components of the system. The \textbf{human-corrected version} further enforces real Mininet/BMv2 execution, corrects lifecycle management and resource cleanup, and achieves successful end-to-end execution on the BMv2 testbed with correct packet exchange and consistent throughput across client and receiver logs.

\item \textbf{Claude Code.}
Across the low, medium, and high settings, Claude Code reconstructs NetChain as a host-side Python artifact rather than as a P4/BMv2 deployment. The lightweight version provides a quick single-script simulator, the medium version improves modularity with dataset loading and functional tests, and the high version shifts toward an analytical or closed-form performance model. These variants can reproduce parts of the experimental workflow or expected performance trends, but they do not generate the full deployable NetChain artifact set.

\item \textbf{Standalone baseline.}
Implements only a hand-written NetChain P4 data plane as a reference artifact, without control plane, client, or orchestration components, and therefore cannot execute end-to-end; it serves purely as a structural ground-truth baseline.
\end{itemize}

\subsubsection{Rateless IBLT}
We evaluate the reproducibility of Rateless IBLT by testing whether different reconstruction strategies, given identical inputs, can generate all required artifacts and successfully execute set-reconciliation experiments on a host-side Python runtime.
\begin{itemize}
\item \textbf{RepLLM.}
Across the automatically generated RepLLM versions, the reconstruction consistently captures the core algorithmic structure of Rateless IBLT, including rateless mapping, encoding, peeling-style decoding, and set-difference reconciliation. The versions differ mainly in input-model fidelity, reconciliation procedure coverage, and whether tests exercise the real protocol path or simplified  components. The \textbf{human-corrected version} replaces validation with tests over the real Rateless IBLT protocol and achieves high consistency with the paper across mapping, encoding, peeling decoding, and set-difference reconciliation.

\item \textbf{Claude Code.}
Across the low, medium, and high settings, Claude Code produces Python Rateless IBLT simulators using the same synthetic dataset. The lightweight version runs end-to-end reconciliation at full scale and includes the Irregular Rateless IBLT variant; the medium version bundles the algorithm, dataset loading, benchmark execution, and report generation into a compact pipeline validated on a smaller sample; and the high version separates the core library from experiment drivers, emphasizing Monte Carlo reproduction of overhead curves and systematic medium-scale benchmarks. Overall, these variants reproduce the main algorithmic behavior but omit some production optimizations and application-level evaluations such as Ethereum state synchronization, multi-peer synchronization, adversarial workloads, SipHash, variable-length counts, and heap-based incremental encoding.

\item \textbf{Standalone baseline.}
The implementation reproduces the paper's core components and correctly supports data loading and encoding. Nevertheless, the current decoder does not maintain a record of recovered items, allowing the same item to be peeled and reinserted multiple times. This causes symbol states to oscillate between pure and non-pure, so that although very small instances terminate successfully, instances with \(d \geq 10\) fail to complete. Consequently, the full evaluation cannot be executed, and the paper's central claims cannot currently be reproduced.
\end{itemize}

\subsubsection{NCFlow}
We evaluate the reproducibility of NCFlow by testing whether different reconstruction strategies, given identical WAN topology and traffic-demand inputs, can generate the required decomposition, routing, and optimization artifacts, and successfully execute multi-commodity flow experiments with third-party solver support.
\begin{itemize}
\item \textbf{RepLLM.}
Across the automatically generated RepLLM versions, the reconstruction progressively covers NCFlow's decomposition, clustering, routing, and optimization pipeline. The generated artifacts vary in how strictly they align with the paper's baseline definitions, input formats, and offline/online execution structure, leading to different levels of agreement with the official implementation. The \textbf{human-corrected version} reaches near-identical total flow compared with the official implementation on the evaluated dataset, while remaining a streamlined single-dataset reproduction that does not cover larger topologies or the full experimental suite in the paper.

\item \textbf{Claude Code.}
Across the low, medium, and high settings, Claude Code implements solver-free simulations on the same dataset. The lightweight version focuses on quickly reproducing the core NCFlow logic on small topologies; the medium version improves modularity and adds metrics such as FIB size while remaining stable on small and medium-scale inputs; and the high version provides a cleaner and more complete structure, although the experiment scale limits reproduction of the strongest speedups and extended scenarios reported in the paper.

\item \textbf{Standalone baseline.}
The standalone baseline lacks proper modular decomposition and orchestration and fails to provide an end-to-end evaluation pipeline.
\end{itemize}

\subsubsection{GRoot}
We evaluate the reproducibility of GRoot by testing whether different reconstruction strategies, given identical inputs (paper, \texttt{metadata.json}, \texttt{jobs.json}, and DNS zone files), can generate the full verification pipeline (zone parser, label graph, equivalence-class generator, symbolic executor, property checkers, and orchestration scripts) and correctly detect all reference violations on the 2 test benchmarks.
\begin{itemize}
\item \textbf{RepLLM.}
Across the automatically generated RepLLM versions, the reconstruction covers the main components of the GRoot verification pipeline, including zone parsing, symbolic execution, equivalence-class construction, interpretation-graph modeling, and property checking. The versions differ in their treatment of equivalence-class refinement, nameserver wiring, wildcard and DNAME semantics, apex handling, and checker alignment. The \textbf{human-corrected version} applies targeted fixes to these semantic and reporting components, aligns all seven \texttt{test1} violation counts with \texttt{output.json} at 165 ECs and 19 JSON entries, and correctly detects the five-hop rewrite chain in \texttt{test2}.

\item \textbf{Claude Code.}
Across the low, medium, and high settings, Claude Code produces host-side Python DNS verifiers with increasing structural completeness. The low version quickly implements zone parsing, equivalence-class generation, and property checking; the medium version modularizes the pipeline and adds layered DNS semantics with broader checker coverage; and the high version introduces formal DNS models, symbolic execution, and more complete property checking. These variants recover important parts of the benchmark behavior, but still show varying degrees of mismatch in equivalence-class counts, violation counts, and entry-level reporting.

\item \textbf{Standalone baseline.}
Provides a hand-written single-file Python DNS verifier with basic zone parsing, simplified equivalence-class generation, and trace-based property checks, but omits full interpretation-graph construction and benchmark-aligned reporting, under-detects most \texttt{test datasets} reference violations and leaves several expected failure categories uncovered.
\end{itemize}

\subsubsection{NetDice}
We evaluate the reproducibility of NetDice by testing whether different reconstruction strategies, given identical inputs (paper, JSON configurations, and topology files), can generate the full analysis pipeline and match reference results on the test dataset.
\begin{itemize}
\item \textbf{RepLLM.}
Across the automatically generated RepLLM versions, the reconstruction covers parsing, routing simulation, property checking, failure exploration, and probabilistic analysis to different degrees. The versions vary in the integration of pipeline stages, hot-edge reasoning, ECMP handling, and bounded exploration. The \textbf{human-corrected version} applies targeted fixes to these components and passes all 11 tests on the same dataset, with probabilities on simple examples, ECMP, and NSFnet benchmarks matching reference values within tolerance.

\item \textbf{Claude Code.}
Across the low, medium, and high settings, Claude Code produces host-side Python probabilistic analyzers over \texttt{test datasets}. The lightweight version implements a single-script pipeline for parsing, BGP simulation, hot-edge identification, and failure exploration; the medium version modularizes dataset loading and batch execution, using exhaustive enumeration on small networks and bounded exploration on larger topologies; and the high version implements NetDice's core algorithms and property types with cold-edge pruning, batch evaluation, and brute-force cross-checks. The high setting achieves the closest agreement with reference probabilities, especially on ECMP and larger-topology benchmarks.

\item \textbf{Standalone baseline.}
A single-file NetDice analyzer covering static routing, independent link failures, and basic property checking on embedded mock input, without BGP simulation, dataset loading, or the full multi-stage pipeline, and therefore cannot support standard end-to-end evaluation; it serves purely as a minimal algorithmic ground-truth baseline.
\end{itemize}

\subsubsection{SelfStarter}
We evaluate the reproducibility of SELFSTARTER by testing whether different implementations can generate the required template-inference artifacts and run structured-generalization experiments from the same router-configuration inputs. SELFSTARTER detects likely misconfigurations by inferring templates for ACLs, prefix lists, and route policies, and identifying structural or parameter outliers across nodes.
\begin{itemize}
\item \textbf{RepLLM.}
Across the automatically generated RepLLM versions, the reconstruction focuses on the ACL template-inference workflow, covering configuration ingestion, block-based structured generalization, line matching, metatemplate construction, group partitioning, parameter mapping, and outlier reporting. The versions progressively improve the organization of the ACL pipeline, including staged execution and Batfish-oriented ingestion, while route-map and prefix-list templating remain outside the main evaluated path. The \textbf{human-corrected version} reproduces the expected evaluation profile on the primary test dataset: seven ACL segments are templated, four are structurally consistent, two group outliers from missing-line structural differences are correctly identified, one parameter-level discrepancy is flagged for manual review, and the final summary reports two group-consistency failures and one parameter-consistency failure, with all other consistency checks passing.

\item \textbf{Claude Code.}
Across the low, medium, and high settings, Claude Code produces Python SELFSTARTER structured-generalization simulators over the same synthetic benchmark. The lightweight version runs end-to-end ACL-template inference on seven ACL configurations from two routers; the medium version integrates Cisco IOS parsing, dataset loading, and benchmark execution into a self-contained pipeline with field-level parameterization; and the high version modularizes parsing, matching, inference, and output generation while adding systematic benchmarking and finer-grained octet-level LGG parameterization. All three focus primarily on ACL structured generalization.

\item \textbf{Standalone baseline.}
The standalone implementation independently reproduces SELFSTARTER as a self-contained system and supports the core structured-generalization workflow for Cisco IOS ACLs. It successfully parses all seven ACLs from two routers in the test1 dataset and reproduces the expected templates, parameter mappings, and outlier classifications. Nevertheless, the implementation relies on a lightweight regular-expression parser and only supports extended ACLs, while routing-policy and prefix-list templating are not implemented. Consequently, test2 cannot be completed because incorrect execution parameters cause segment matching to return an empty result.
\end{itemize}

\subsubsection{Teal}
We evaluate the end-to-end reproducibility of Teal by testing whether different implementations can train and evaluate path-formulation traffic allocations from the same WAN topologies, traffic matrices, and candidate-path inputs. Teal accelerates wide-area traffic engineering by combining a flow-centric graph neural network to encode topology and demand features, multi-agent reinforcement learning to allocate each traffic demand toward a global objective, and ADMM fine-tuning to reduce link-capacity violations across nodes.
\begin{itemize}
\item \textbf{RepLLM.}
Across the automatically generated RepLLM versions, the reconstruction establishes Teal's main train-and-evaluate framework, including topology and traffic-matrix ingestion, candidate-path handling, FlowGNN-style encoding, COMA-style policy training, ADMM-based post-processing, metric aggregation, logging, checkpoints, and allocation export. The versions differ in the completeness of model integration, graph/tensor conversion, inference consistency, ADMM formulation, and large-topology scalability. The \textbf{human-corrected version} reproduces the expected evaluation profile on the primary datasets: on \textbf{test/B4}, it loads real traffic matrices and cached candidate paths, trains the FlowGNN--COMA--ADMM stack end-to-end, outputs non-negative split ratios with per-demand sums not exceeding one, and reports total flow, satisfied-demand ratio, maximum link utilization, capacity violations, and sub-second per-interval runtime, with measurable ADMM improvement over the policy-only allocation; on \textbf{test/UsCarrier}, it completes training and evaluation with the larger-topology ADMM iteration budget when path caches are available. Kdl and ASN2k remain outside the default audit path because absent path caches make on-the-fly generation infeasible on CPU, and full-scale training on these topologies requires substantially more GPU memory than the lightweight smoke setting provides.

\item \textbf{Claude Code.}
Across the low, medium, and high settings, Claude Code produces Python Teal traffic-allocation simulators implementing FlowGNN-style encoding, multi-agent reinforcement learning, and ADMM fine-tuning over the same WAN traffic-engineering datasets. The lightweight version completes end-to-end training and evaluation on B4 with full COMA policy-gradient training and ADMM; the medium version bundles data loading, simplified COMA training, ADMM, and report generation into a single self-contained module; and the high version separates the core library from experiment drivers, emphasizing per-interval benchmarks, ADMM ablations, structured metric export, and UsCarrier forward-pass checks. These variants reproduce the B4 workflow most completely, while large-scale end-to-end training, LP/NCFlow baselines, link-failure experiments, and GPU speedups remain outside the reproduced scope.

\item \textbf{Standalone baseline.}
Standalone attempts to reproduce Teal's high-level framework, including path-based traffic engineering, candidate-path generation, FlowGNN, COMA-based training, and simplified ADMM fine-tuning. However, the implementation cannot complete dataset loading or execution due to data-format assumptions, hard-coded dimensional errors, and indexing bugs, so even the limited B4 evaluation cannot be reproduced, and the paper's scalability and performance claims remain unsupported.
\end{itemize}


\subsection{Prompts of LLM-based Evaluation}
We provide the prompts for LLM-based global-level semantic alignment evaluation in Figure~\ref{fig:llm_as_a_judge_total_1}.
\begin{figure*}[ht]
\centering
  \includegraphics[width=0.85\linewidth]{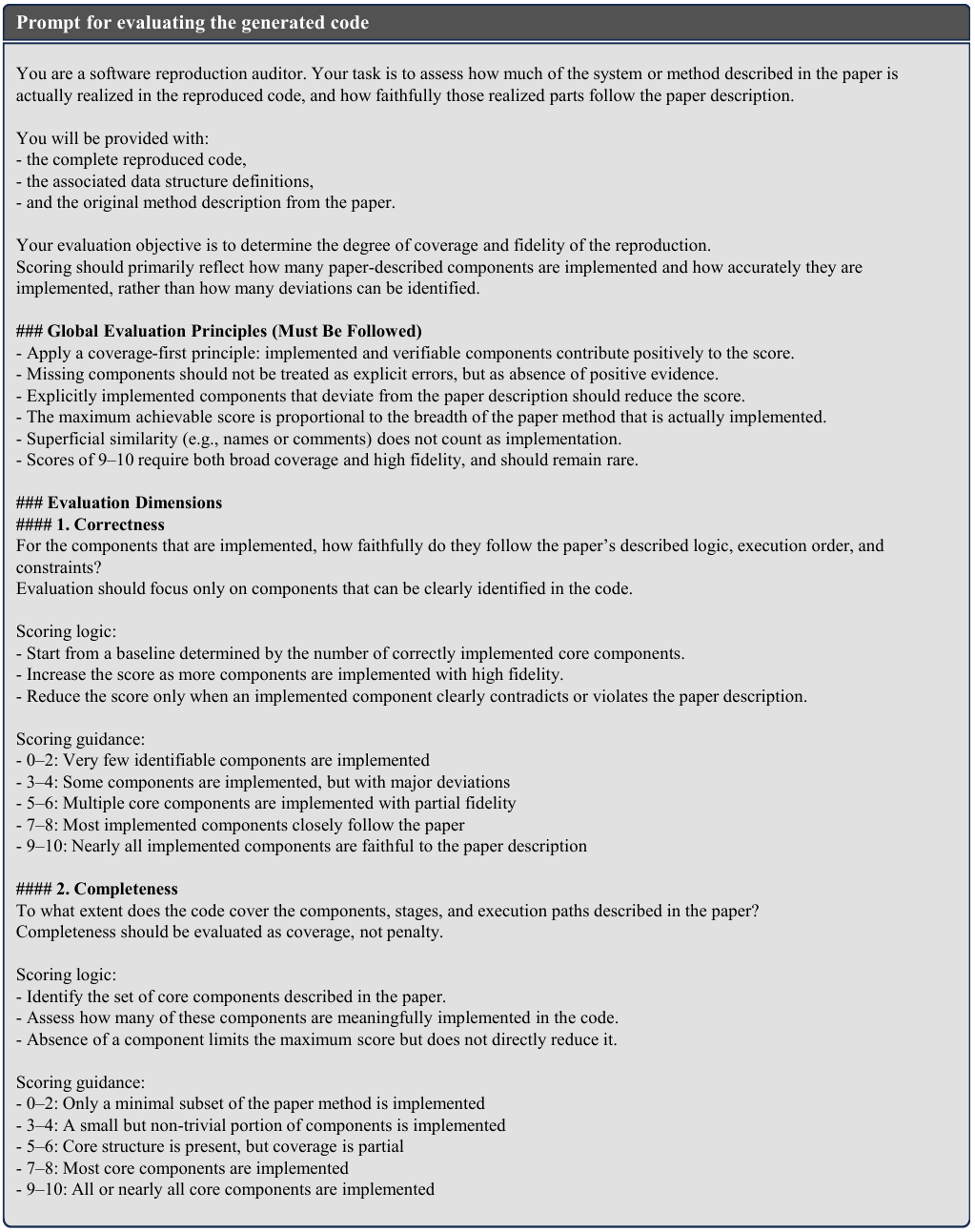}
  \caption{Prompts for LLM-based Paper–Code Semantic Alignment Evaluation}
  \label{fig:llm_as_a_judge_total_1}
\end{figure*}
\begin{figure*}[ht]
\centering
  \includegraphics[width=0.9\linewidth]{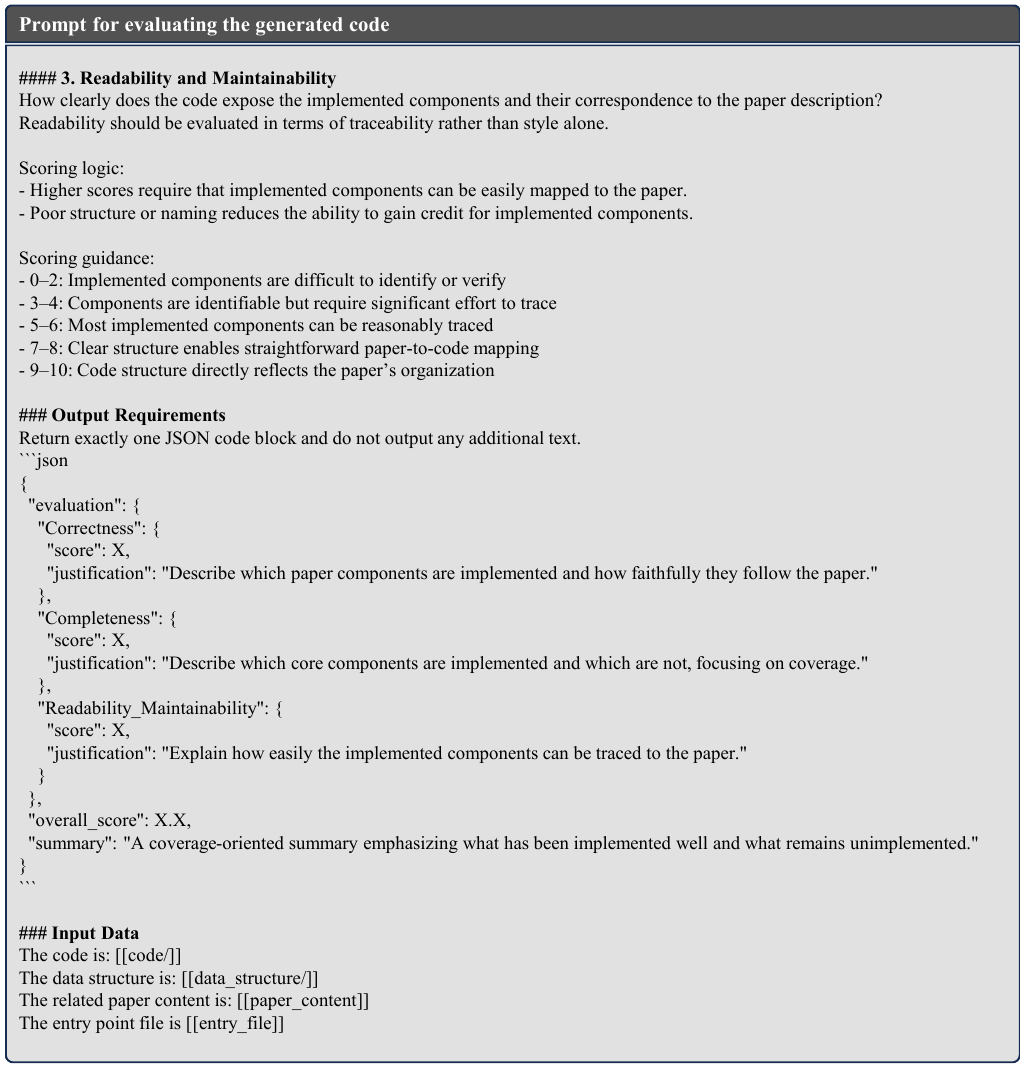}
  \label{fig:llm_as_a_judge_total_2}
\end{figure*}

\end{document}